\documentclass[aps,prl,twocolumn,amsmath,amssymb,superscriptaddress]{revtex4-1}

\usepackage{graphics}
\usepackage{dcolumn}
\usepackage{bm}
\usepackage[caption=false]{subfig} 
\usepackage{color}
\usepackage{tikz}
\usepackage{subfig}
\usepackage[titles,subfigure]{tocloft}
\usepackage{soul}

\usepackage[titletoc,toc,title]{appendix}

\begin{document}

\title{Projected BCS Theory for the Unification of Antiferromagnetism and Strongly Correlated Superconductivity}

\author{Hyunwoong Kwon}
\affiliation{Quantum Universe Center, Korea Institute for Advanced Study, Seoul 02455, Korea}
\author{Kwon Park}
\email[Electronic address:$~~$]{kpark@kias.re.kr}
\affiliation{Quantum Universe Center, Korea Institute for Advanced Study, Seoul 02455, Korea}
\affiliation{School of Physics, Korea Institute for Advanced Study, Seoul 02455, Korea}
\affiliation{Department of Physics, Harvard University, Cambridge, Massachusetts 02138, USA}
\date{\today}

\begin{abstract}
The intimate connection between antiferromagnetism and superconductivity is at the core of high-temperature superconductivity.
Here, we put forward the projected BCS theory for the unification of antiferromagnetism at half filling and strongly correlated superconductivity at moderate doping.
Specifically, it is shown that the projected BCS theory provides excellent trial states for the exact ground states of the $t$-$J$ model in the square lattice, generating the unified phase diagram as a continuous function of hole concentration.
Precisely capturing antiferromagnetism at half filling, which is ultimately a consequence of the strong correlation between Cooper pairs, the projected BCS theory is able to produce better trial states for strongly correlated superconductivity at moderate doping than the resonating valence bond state.
Finally, we discuss various ramifications of the projected BCS theory.
\end{abstract}

\maketitle


Despite the large variety in material properties, there is a certain list of common features robust across high-temperature superconductors. 
One of the most salient features in such a list is the proximity between antiferromagnetism and superconductivity. 
An important question is exactly how these two phenomena are connected together.

The spin singlet is the smallest unit of antiferromagnetism. 
Since the BCS state is the condensate of spin-singlet electron pairs, i.e., Cooper pairs, it is rather natural that antiferromagnetism, at least with short-range order, is closely connected with superconductivity.  
The resonating valence bond (RVB) state is the trial state constructed under this very rationale. 
Specifically, the RVB state is the projected BCS state with all the components containing doubly occupied sites projected out~\cite{Anderson87, Baskaran87, Gros88, Yokoyama88, Paramekanti01, Sorella02, Paramekanti04, Anderson04}. 
Known as the Gutzwiller projection when fully applied, this projection can be also partially imposed~\cite{Bernevig03, Zhang03, Misawa14}. 

A problem is that the RVB state reduces to the spin liquid with fluctuating spin singlets rather than the long-range-ordered antiferromagnet at half filling.
Quantum antiferromagnets in two dimensions can, in principle, have various types of spin liquids and valence bond solids~\cite{Sachdev02}.
However, does this mean that antiferromagnetism with true long-range order is necessarily incompatible with superconductivity?

Here, we show that it is possible to construct a unified theory of antiferromagnetism (AF) and strongly correlated superconductivity (SCSC).
A gist of the unified theory is that the Gutzwiller projection, not commuting with the BCS Hamiltonian, should be treated better.
Specifically, in this theory, the Gutzwiller projection is applied onto the BCS Hamiltonian itself: the BCS Hamiltonian is directly diagonalized in the Gutzwiller-projected Hilbert space~\cite{Park05_PRL, Park05_PRB}. 
For convenience, let us call this theory the projected BCS theory.

To provide the validity of the projected BCS theory, we perform exact diagonalization of the projected BCS Hamiltonian and compare the so-obtained exact ground state with that of the $t$-$J$ model.
Actually, the projected BCS theory has been previously analyzed by using the similar numerical technique~\cite{Park05_PRL, Park05_PRB}. 
While providing promising results, however, the previous analysis has suffered from finite-size effects. 
Here, we overcome this problem by devising the proper overlap in the presence of particle number fluctuations. 

As a result, it is shown that the projected BCS theory provides excellent trial states for the exact ground states of the $t$-$J$ model in the square lattice for a wide range of hole concentration. 
In particular, we provide a rigorous proof for the equivalence between the exact ground states of the $t$-$J$ model and the projected BCS theory at half filling.
More importantly, we obtain the unified phase diagram of the $t$-$J$ model as a continuous function of hole concentration, capturing both AF at half filling and SCSC at moderate doping.
The Fermi liquid state is also naturally captured at sufficiently large doping.
This phase diagram provides concrete evidence for the split between the pairing amplitude and the real superconducting order parameter at low doping, which can in principle explain the pseudogap phenomenon. 

It is emphasized that, precisely capturing AF at half filling, the projected BCS theory is able to produce better trial states for SCSC at moderate doping than the RVB state, demonstrating the merit of the unification of AF and SCSC, as advocated by the $SO(5)$ theory~\cite{Demler04}.
Finally, the projected BCS theory can be used to reveal the relevance of the $s$-wave pairing symmetry as well as the geometrical frustration to high-temperature superconductivity.

{\bf Results}

{\bf High-temperature superconductivity and the fractional quantum Hall effect.}
There is a remarkable similarity between the RVB state for high-temperature superconductivity and the composite fermion (CF) state~\cite{Jain89} for the fractional quantum Hall effect (FQHE). 
Simply put, the RVB state is the BCS state projected onto the Hilbert space with no double occupancy:
\begin{align}
\psi_{\rm RVB} ={\cal P}_G \psi_{\rm BCS},
\end{align}
where ${\cal P}_G$ is the Gutzwiller projection operator, and $\psi_{\rm BCS}$ is the BCS wave function describing the electron-electron paired state composed of Cooper pairs. 
Similarly, the CF state is the electron-vortex bound state projected onto the lowest Landau level (LLL): 
\begin{align}
\psi_{\rm CF}={\cal P}_{\rm LLL}  \psi_{\rm Jastrow},
\end{align}
where ${\cal P}_{\rm LLL}$ is the LLL projection operator, and $\psi_{\rm Jastrow}$ is the Jastrow-correlated wave function describing the electron-vortex bound state composed of CFs~\cite{Jain89}.
The electron-electron pairing in the RVB state is a consequence of the intricate interplay between the no-double-occupancy constraint and the residual hopping effect. 
Similarly, the electron-vortex binding in the CF state is a consequence of that between the LLL constraint and the residual electron-electron interaction. 

One of the most important and convenient properties of these two states is that both Cooper pairs and CFs themselves can be treated as weakly interacting.  
The `plain vanilla' version of the RVB theory assuming essentially non-interacting Cooper pairs can provide a semi-quantitative understanding of high-temperature superconductivity~\cite{Anderson04}. 
Similarly, the FQHE can be understood as the integer quantum Hall effect of weakly interacting CFs~\cite{Jain89}.
The trial wave function based on weakly interacting CFs has been incredibly successful, basically explaining all known properties of the FQHE~\cite{Jain_Book}.  

The similarity between high-temperature superconductivity and the FQHE goes even deeper in that there are fundamental difficulties in the field theoretical implementation of the required projection operators, ${\cal P}_G$ and ${\cal P}_{\rm LLL}$, respectively. 
Specifically, various gauge theories, often in terms of slave particle methods, have been widely used to implement ${\cal P}_G$~\cite{Lee06}.
Similarly, the Chern-Simons gauge theory has provided a reasonable low-energy description of the FQHE (and also the CF sea state) by implementing ${\cal P}_{\rm LLL}$ in the mean-field and one-loop levels~\cite{Lopez91, Kalmeyer92, HLR93}.
Despite these successes, however, it is necessary to treat the projection operators better to capture the effects of strong correlation more accurately.

There are several important examples in the FQHE, where the correlation between CFs cannot be ignored.
The first example is the Wigner crystal occurring at low filling factors in the LLL~\cite{Jiang90}, where CFs, rather than electrons, form the Wigner crystal~\cite{Yi98, Archer13}.
The second example is the famous $5/2$ state, the even-denominator FQHE state occurring in the half-filled second Landau level~\cite{Willett87}.
Let us elaborate on the second example below.

It is generally believed that the $5/2$ state is a paired quantum Hall state~\cite{MooreRead91, Greiter92, Park98, Scarola00, Lu10}. 
Despite this general belief, however, there are various crucial questions to be answered.
Among such questions are the proof of pairing other than the gap, the pairing symmetry, the validity of the non-Abelian braiding statistics, and so on.
Considering the effectiveness of the trial state method in the FQHE, the better trial state can be constructed, the more conclusive answers can be obtained.

One of the most promising trial states for the $5/2$ state is the Moore-Read (MR) Pfaffian state~\cite{MooreRead91, Greiter92} along with its particle-hole (PH) conjugate, called the anti-Pfaffian state, and their PH-symmetrized combination~\cite{RezayiHaldane00, Peterson08}. 
While the MR Pfaffian state can be written explicitly in the planar geometry, it is generally more convenient to construct it by directly diagonalizing the model Hamiltonian, whose exact ground state is known to be the MR Pfaffian state.
Such a model Hamiltonian is the three-body repulsive interaction, often denoted as $H_3$, imposing an energy penalty to closely-packed three-electron clusters. 
Consequently, checking the validity of the MR Pfaffian state amounts to computing the overlap between the exact ground states of the Coulomb interaction and $H_3$~\cite{RezayiHaldane00, Peterson08}. 

By the same token, to capture the long-range antiferromagnetic order at half filling, one needs to better take into account the strong correlation between Cooper pairs. 
To this end, we consider the projected BCS Hamiltonian.

{\bf Projected BCS Hamiltonian.}
The projected BCS Hamiltonian can be written as follows:
\begin{align}
H_{\rm PBCS} = {\cal P}_G ( H_t +H_\Delta +H_\mu ) {\cal P}_G,
\label{eq:H_PBCS}
\end{align}
where
\begin{align}
H_t &= -t \sum_{\langle i,j \rangle, \sigma} ( c^\dagger_{i\sigma} c_{j\sigma} +{\rm H.c.} ) , \nonumber \\
H_\Delta &=\sum_{\langle i,j \rangle} \Delta_{ij} ( c^\dagger_{i\uparrow} c^\dagger_{j\downarrow} -c^\dagger_{i\downarrow} c^\dagger_{j\uparrow}+{\rm H.c.} ) , \nonumber \\
H_\mu & = -\mu \sum_i n_i ,
\label{eq:H_PBCS_detailed}
\end{align}
where $t$ is the hopping parameter, $\Delta_{ij}$ is the pairing amplitude, and $\mu$ is the chemical potential.
For the $d$-wave pairing symmetry, $\Delta_{ij}=\pm\Delta$ for the $x$ and $y$ directions, respectively.
For the $s$-wave pairing symmetry, $\Delta_{ij}=\Delta$ for both directions.
The particle number operator $n_i$ is given by $n_i=\sum_\sigma c^\dagger_{i\sigma}c_{i\sigma}$.

In the meantime, the $t$-$J$ Hamiltonian can be written as follows:
\begin{align}
H_{t\mbox{-}J} = {\cal P}_G ( H_t +H_J ) {\cal P}_G,
\end{align}
where $H_t$ is the same as in Eq.~\eqref{eq:H_PBCS_detailed}, and
\begin{align}
H_J &= J \sum_{\langle i,j \rangle} \left( {\bf S}_i \cdot {\bf S}_j -n_i n_j/4  \right) ,
\end{align}
where $J$ is the spin exchange coupling constant with $J>0$~\cite{Gros87, MacDonald88, Zhang88}.
Note that, unless mentioned otherwise, all the energy scales are measured in units of $t$ throughout this work.

Similar to what is done for the $5/2$ state, we would like to compute the overlap between the exact ground states of the $t$-$J$ model and the projected BCS theory.  
Before performing actual computations, however, it is important to check the symmetries of the Hamiltonians, which can crucially affect the overlap. 
To this end, we perform group theoretical analyses, whose details are provided in {\bf Methods}.

More importantly, there is a certain limit, where the value of the overlap is exactly known. 
That is, the overlap should be exactly unity at half filling. 

{\bf Equivalence at half filling.}
The equivalence between the $t$-$J$ model and the projected BCS theory at half filling can be proved analytically.
With the details of the proof presented in {\bf Methods}, here, we provide an overall summary.

We begin by decomposing the Schr\"{o}dinger equation of the projected BCS theory in terms of a system of simultaneous equations connecting between different number sectors. 
Then, we take the limit of large $\mu$ to maximize the electron number under the no-double-occupancy constraint, which leads to half filling.    
In this limit, we carefully expand the weighting amplitude of the energy eigenstate in each particle number sector in orders of $1/\mu$.
Finally, by keeping only the most dominant terms, it is shown that the projected BCS theory becomes entirely equivalent to the Heisenberg model, i.e., the $t$-$J$ model at half filling.

Several aspects of the proof are worthwhile to mention.
First, the equivalence is not just for the ground state, but rather the entire energy eigenstates at half filling.
Second, the proof works regardless of the pairing symmetry or the lattice structure.
As shown by actual computations, the overlap is exactly unity at half filling not only for the $d$-wave pairing symmetry, but also for the $s$-wave pairing symmetry in the square lattice.
The same is true for the triangular lattice.
Third, the equivalence at half filling does not necessarily guarantee a high overlap upon doping.
The overlap at finite doping depends crucially on the pairing symmetry and the lattice structure. 
It is shown that the overlap can remain high with finite $\Delta$ for the $d$-wave pairing symmetry in the square lattice, but not for the $s$-wave pairing symmetry. 
The overlap is generally quite low at moderate doping in the triangular lattice.

To actually compute the overlap between the exact ground states of the $t$-$J$ model and the projected BCS theory, now, let us consider the proper overlap in the presence of particle number fluctuations.

\begin{figure*}
\centering
\includegraphics[width=1.7\columnwidth]{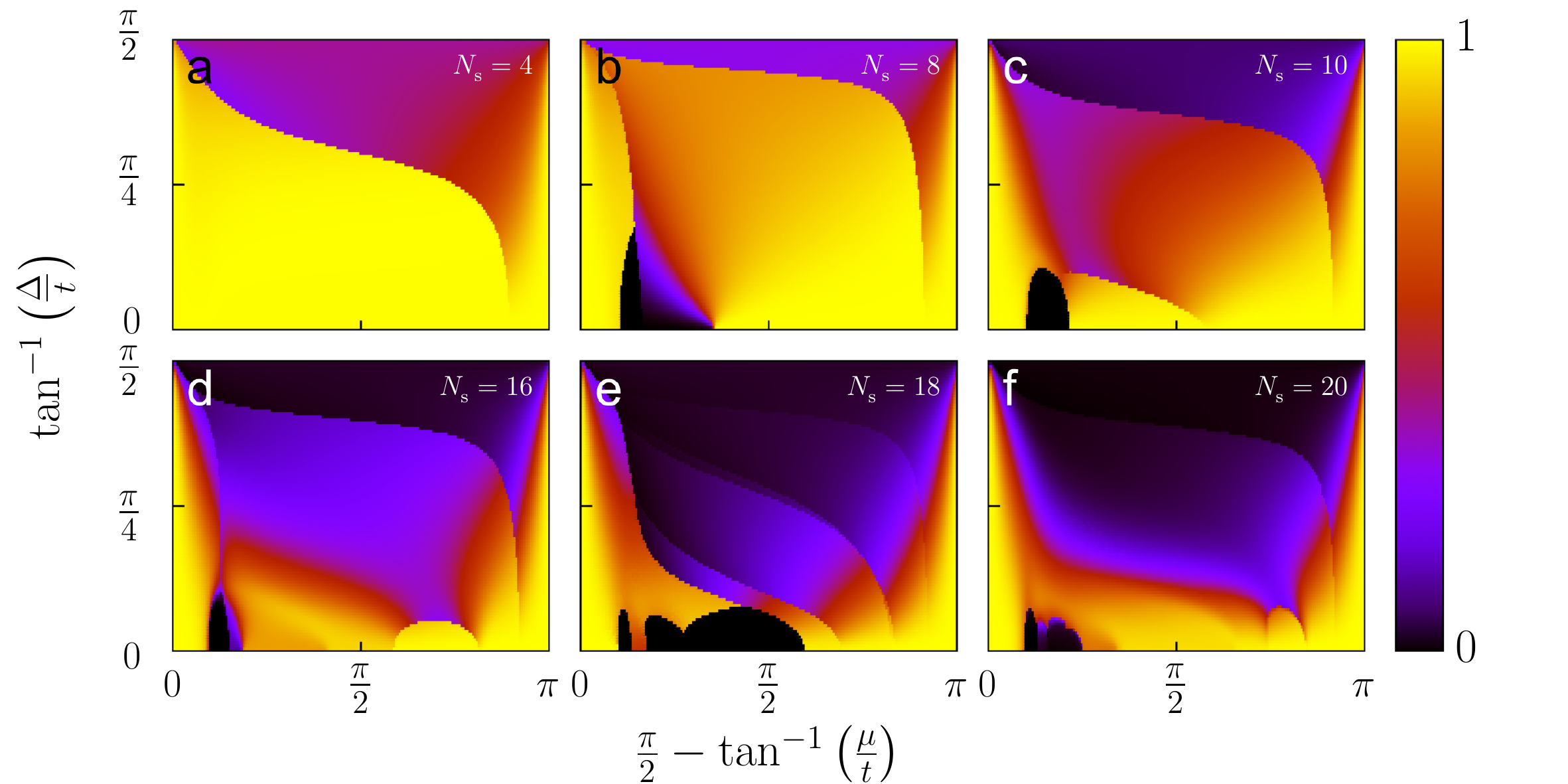}
\caption{{\bf Phase diagram of the $t$-$J$ model in the square lattice determined via the number-weighted overlap squared (NWOS) between the exact ground states of the $t$-$J$ model and the projected BCS theory with the $d$-wave pairing symmetry.}
The spin exchange coupling constant of the $t$-$J$ model is set to be $J/t=0.5$.
Denoted in color with bright yellow being unity and black being zero, the NWOS is computed as a function of pairing amplitude $\Delta$ and chemical potential $\mu$.
As one can see, the NWOS is exactly unity regardless of $\Delta$ at sufficiently large $\mu$, proving that the long-range antiferromegnetic order is precisely captured by the projected BCS theory. 
Note that the NWOS also becomes unity as $\mu$ approaches $-\infty$, i.e., the vacuum, where both spin exchange and pairing terms play negligible roles.
The number of sites is varied as $N_s = 4, 8, 10, 16, 18, 20$.
}
\label{fig:PBCSmu}
\end{figure*}

\begin{figure*}
\centering
\includegraphics[width=1.7\columnwidth]{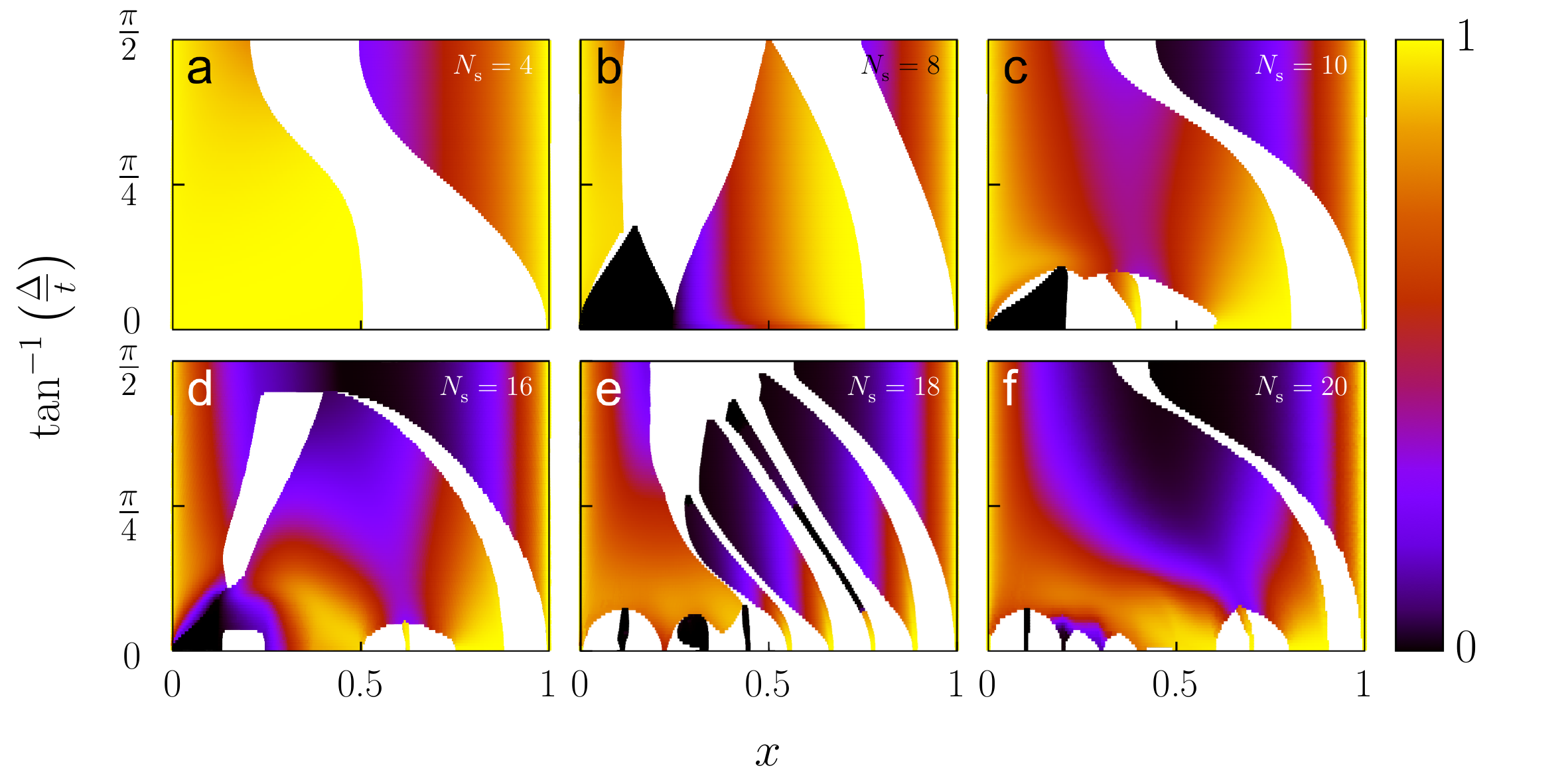}
\caption{{\bf Similar phase diagram as Fig.~\ref{fig:PBCSmu} with the abscissa converted to the hole concentration $x$.}
$\mu$ is related with $x$ via $x = 1-\sum_i \langle n_i \rangle /N_s$.
The optimal pairing amplitude producing the maximum NWOS, $\Delta_{\rm max}$, can be determined as a function of $x$.
Note that the NWOS is exactly unity regardless of $\Delta$ at $x=0$, i.e., half filling, and $x=1$, i.e., the vacuum.
The white, or vacant regions indicate that no ground states are available there for the projected BCS theory.}
\label{fig:PBCSx}
\end{figure*}

\begin{figure}
\centering
\includegraphics[width=0.7\columnwidth]{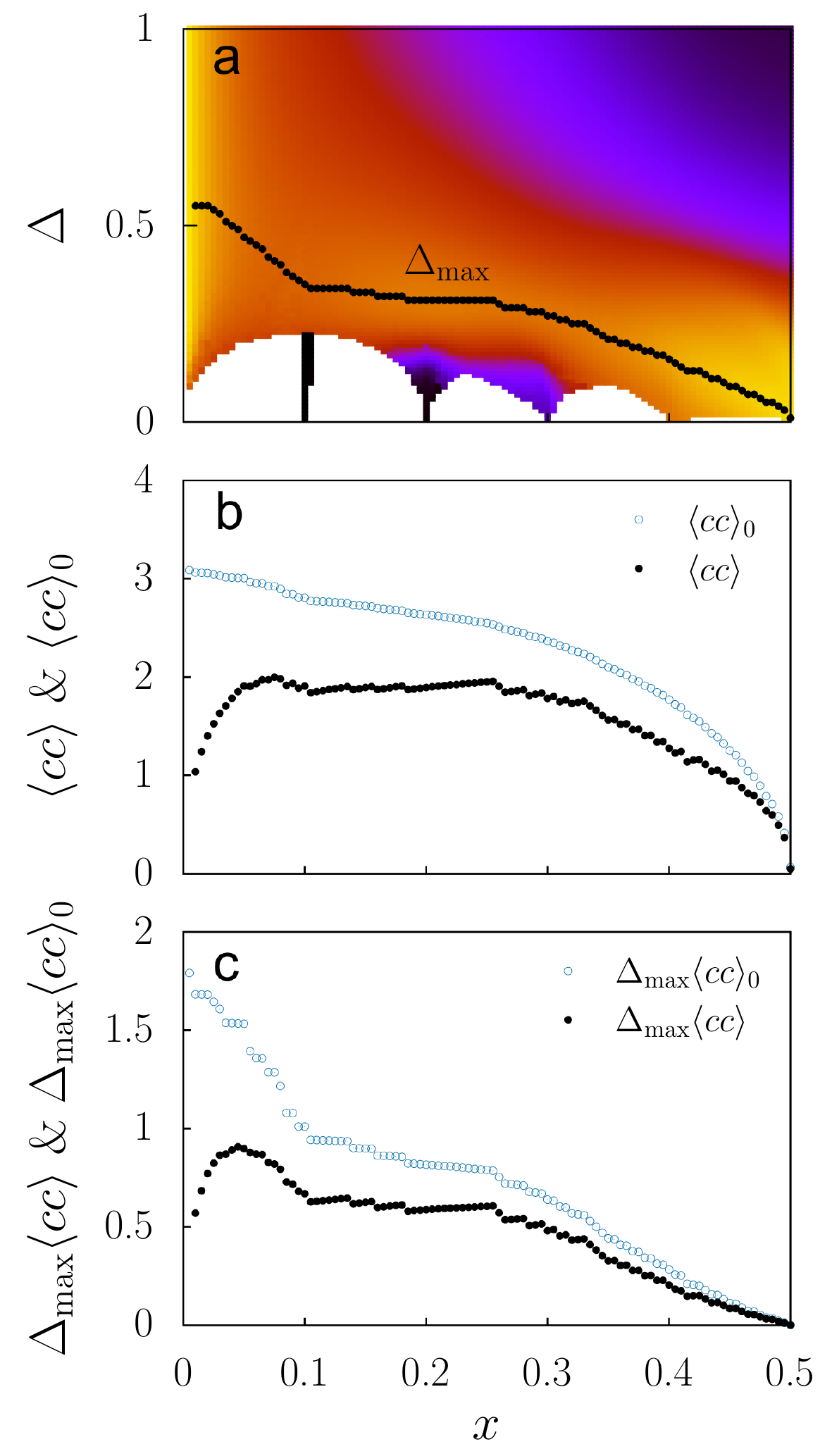}
\caption{{\bf Optimal pairing amplitude $\Delta_{\rm max}$ and superconducting order parameter $\langle cc \rangle$ for the projected BCS theory with the $d$-wave pairing symmetry in the square lattice.}
({\bf a}) 
$\Delta_{\rm max}$ is plotted as solid circles denoting the locations of the maximum NWOS as a function of $x$.
({\bf b}) The superconducting order parameter $\langle cc \rangle$ is computed by using the exact ground states of the projected BCS theory with $\Delta_{\rm max}$ as an input parameter.
$\langle cc \rangle$ is the abbreviation of the superconducting order parameter $\langle c_{i\uparrow} c_{j\downarrow} \rangle$ with $i$ and $j$ being the nearest neighbors.
Note that the sign of $\langle c_{i\uparrow} c_{j\downarrow} \rangle$ depends on whether $j$ is the nearest neighbor of $i$ along the $x$ or $y$ direction, as set by the $d$-wave pairing symmetry.
Similarly abbreviated, $\langle cc \rangle_0$ indicates the ``bare'' superconducting order parameter.
({\bf c}) $\Delta_{\rm max} \langle cc \rangle$ provides the expectation value of the pairing energy with $\Delta_{\rm max} \langle cc \rangle_0$ being its bare counterpart.
These results are obtained at $N_s = 20$.} 
\label{fig:SC_order_parameter}
\end{figure}

\begin{figure*}
\centering
\includegraphics[width=1.7\columnwidth]{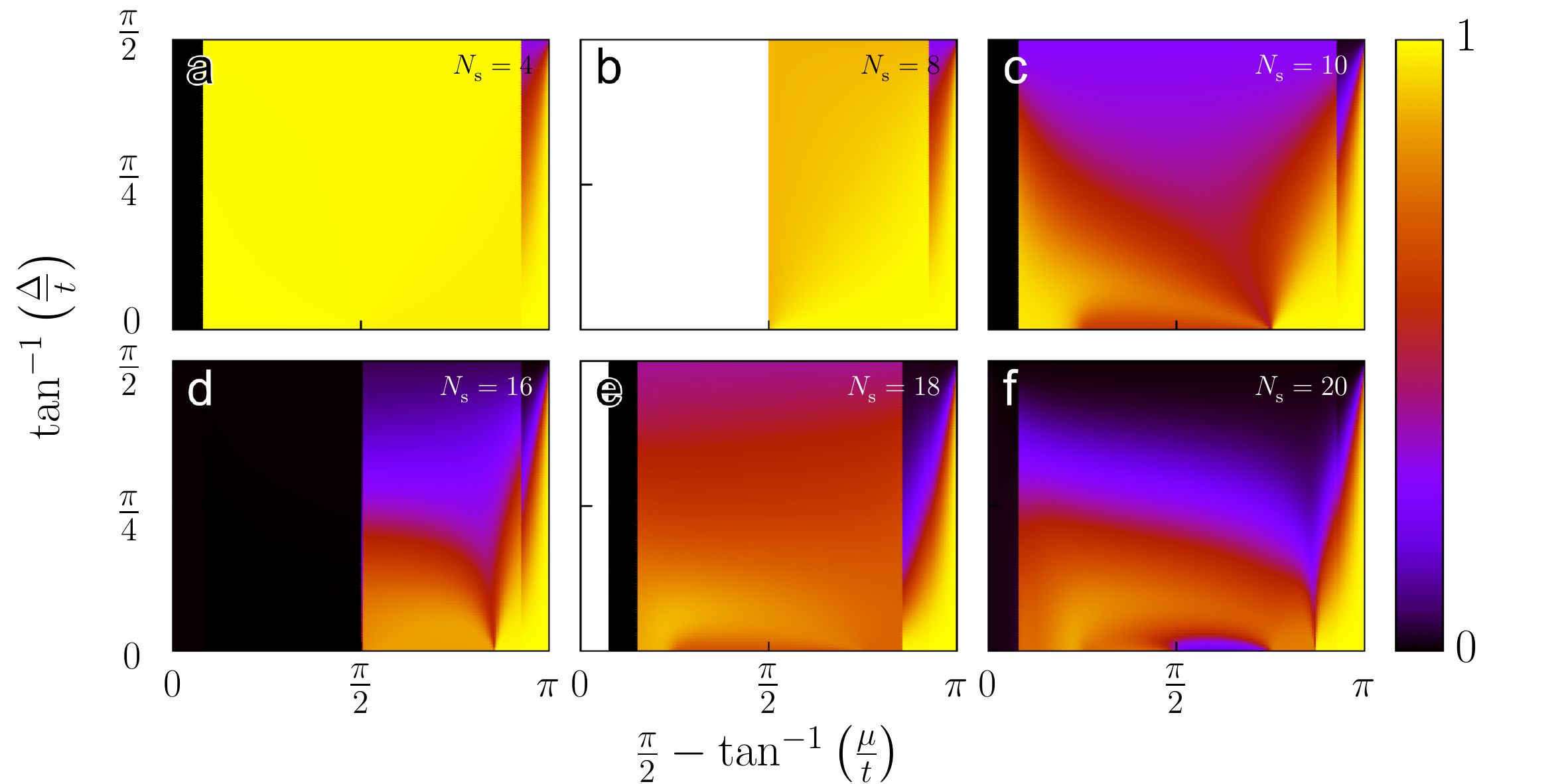}
\caption{{\bf Phase diagram of the $t$-$J$ model in the square lattice determined via the NWOS between the exact ground state of the $t$-$J$ model and the RVB state with the $d$-wave pairing symmetry.}
Similar to Fig.~\ref{fig:PBCSmu}, the spin exchange coupling constant of the $t$-$J$ model is set to be $J/t=0.5$.
The white regions indicate that no RVB states are available there.
Note that the NWOS is entirely zero regardless of $\Delta$ at sufficiently large $\mu$ as expected from the fact that the RVB state cannot capture the long-range antiferromagnetic order. 
}
\label{fig:RVBmu}
\end{figure*}

\begin{figure*}
\centering
\includegraphics[width=1.7\columnwidth]{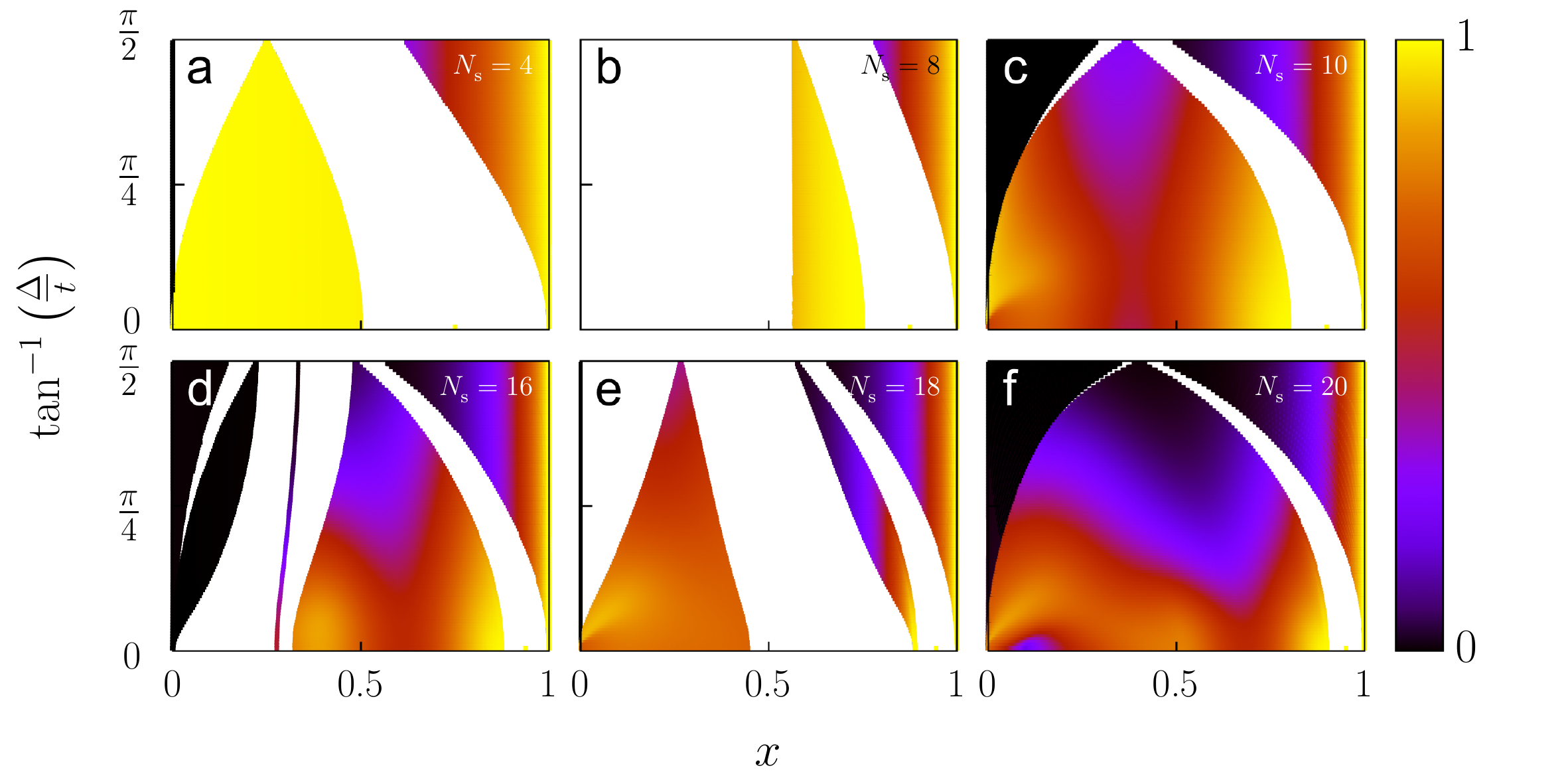}
\caption{{\bf Similar phase diagram as Fig.~\ref{fig:RVBmu} with the abscissa converted to the hole concentration $x$.}
The white regions indicate that no RVB states are available there.
Note that the NWOS is entirely zero strictly at half filling.
}
\label{fig:RVBx}
\end{figure*}

\begin{figure}
\centering
\includegraphics[width=0.7\columnwidth]{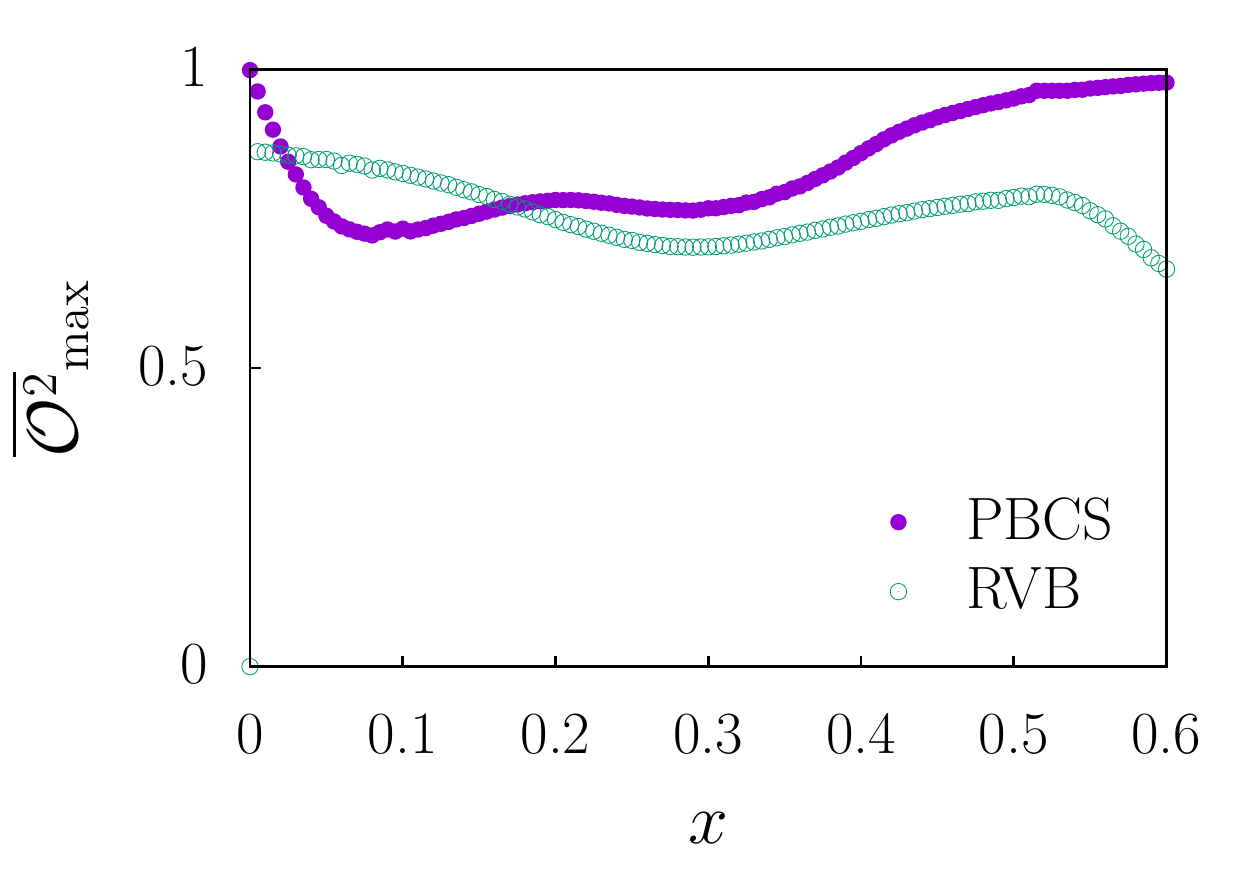}
\caption{{\bf Maximum NWOS, $\overline{{\cal O}^2}_{\rm max}$, of the projected BCS theory in comparison with that of the RVB state as a function of $x$.}
Note that these results are obtained at $N_s=20$. 
}
\label{fig:NWOS_max}
\end{figure}

\begin{figure*}
\centering
\includegraphics[width=1.7\columnwidth]{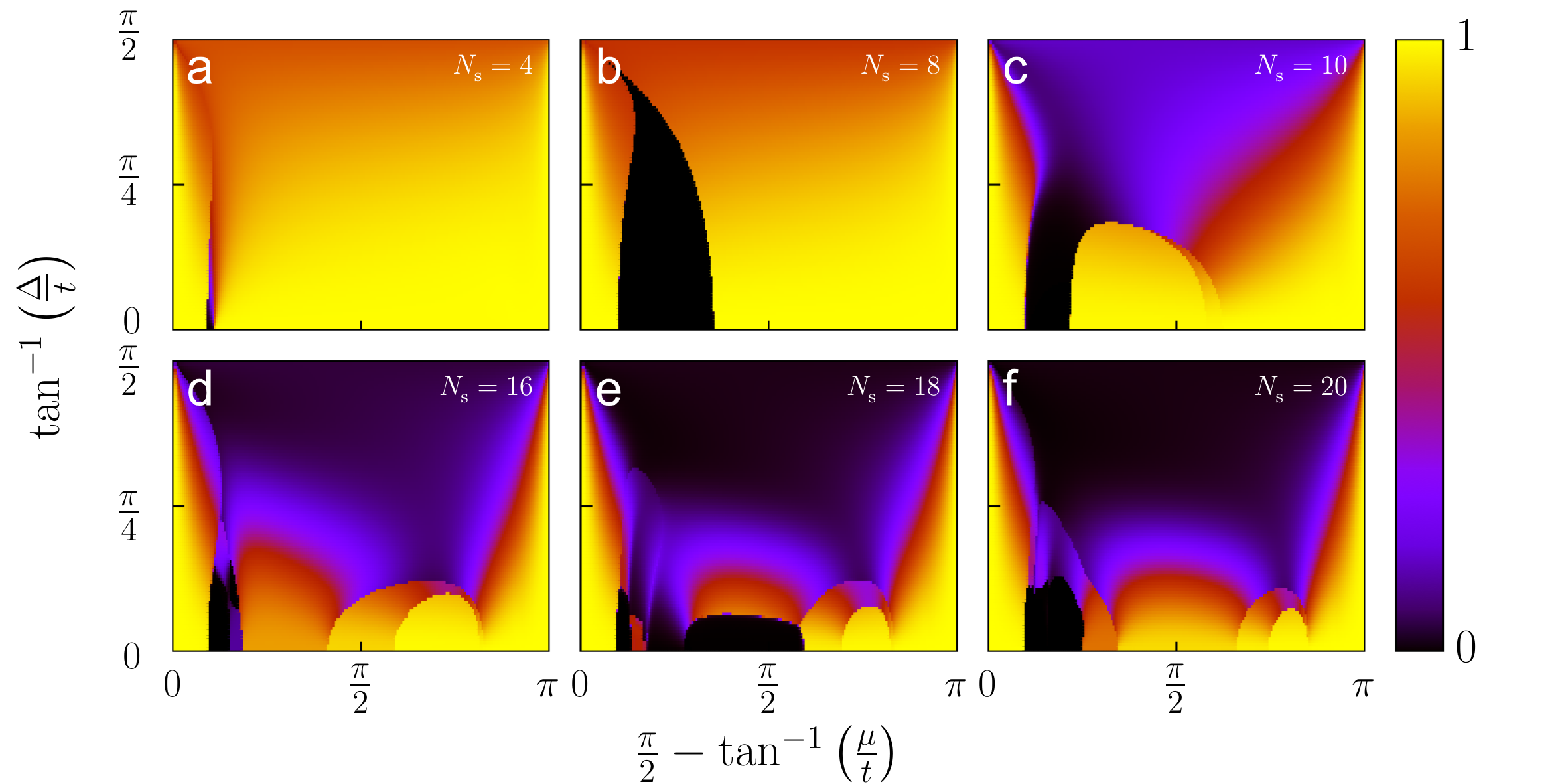}
\caption{{\bf Phase diagram of the $t$-$J$ model in the square lattice determined via the NWOS between the exact ground states of the $t$-$J$ model and the projected BCS theory with the $s$-wave pairing symmetry.}
Similar to Fig.~\ref{fig:PBCSmu}, the spin exchange coupling constant of the $t$-$J$ model is set to be $J/t=0.5$.
As one can see, the NWOS is always maximized along $\Delta=0$ for all finite $\mu$ with a seeming exception at $N_s=18$. 
Note that, at $N_s=18$, the region of otherwise maximized NWOS at $\Delta=0$ is simply masked by the black dome around $\mu=0$.
}
\label{fig:s-wave}
\end{figure*}

\begin{figure}
\centering
\includegraphics[width=0.7\columnwidth]{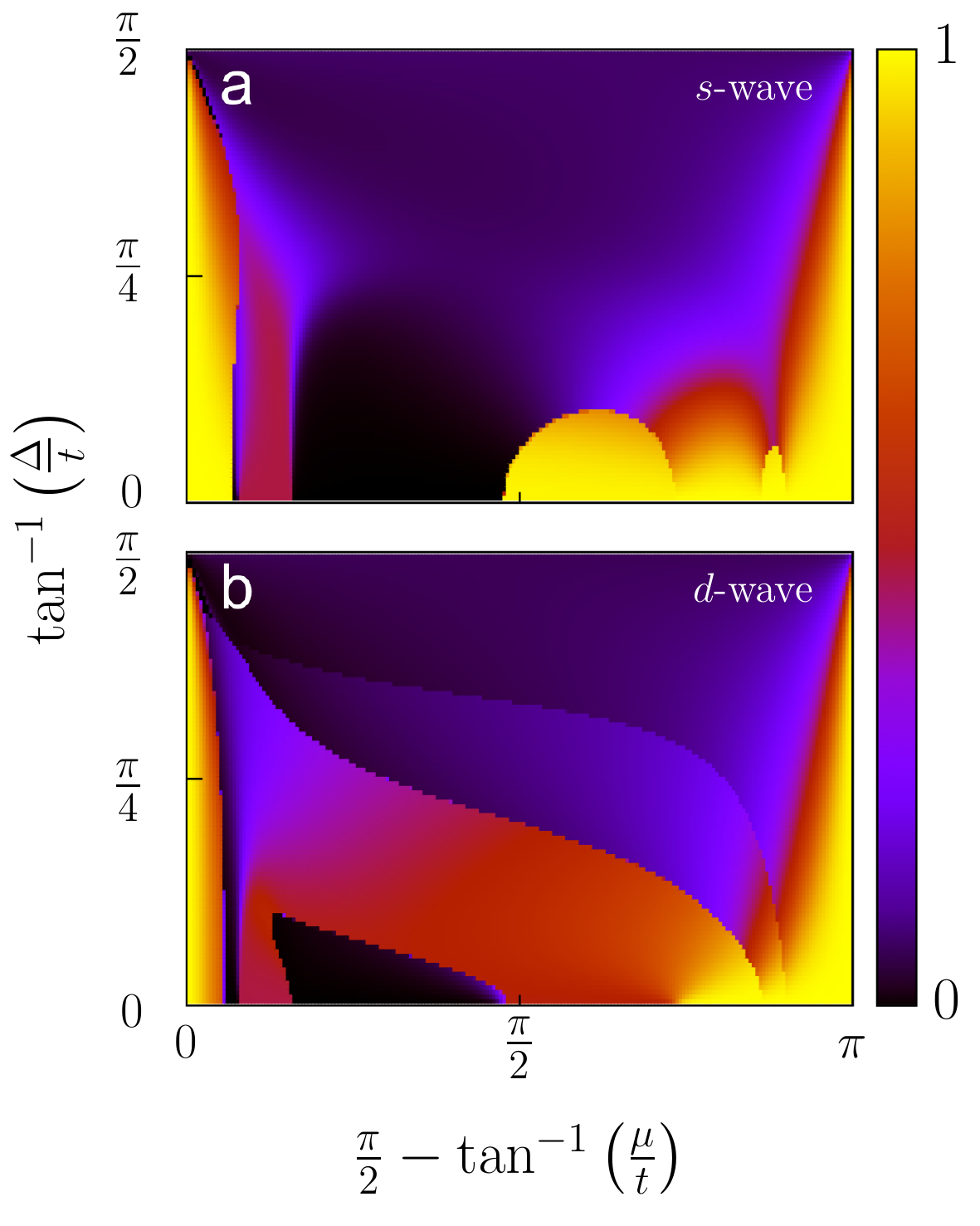}
\caption{{\bf Phase diagram of the $t$-$J$ model in the triangular lattice determined via the NWOS between the exact ground states of the $t$-$J$ model and the projected BCS theory.}
The pairing symmetries are chosen to be $s$- and $d$-wave in ({\bf a}) and ({\bf b}), respectively.
Note that this result is obtained in the $N_s=12$ system, which is the only triangular-lattice system accessible via exact diagonalization, satisfying correct symmetries. 
}
\label{fig:Triangular}
\end{figure}

{\bf Proper overlap in the presence of particle number fluctuations.}
The exact ground state of the projected BCS theory can be expanded in terms of its components in various particle number sectors:
\begin{align}
|\psi_{\rm PBCS}\rangle = \sum_{h=0}^{N_s} \lambda_h |\phi^{\rm PBCS}_h\rangle,
\label{eq:hole_expansion}
\end{align}
where $h$ is the number of holes, related with those of sites, $N_s$, and electrons, $N_e$, via $h=N_s-N_e$.
The RVB state can be also similarly expanded.

Meanwhile, any exact ground state of the $t$-$J$ model has a fixed number of particles. 
This means that each particle number sector has its own exact ground state of the $t$-$J$ model, forming a set of $\{|\phi^{t\mbox{-}J}_h\rangle\}$.
Now, the question is how to compute the overlap between $|\psi_{\rm PBCS}\rangle$ and $\{|\phi^{t\mbox{-}J}_h\rangle\}$. 

A traditional method is to project $|\psi_{\rm PBCS}\rangle$ onto  a specific particle number sector and compute its overlap with $|\phi^{t\mbox{-}J}_h\rangle$ in that sector, i.e.,
\begin{align}
{\cal O}^2_h=|\langle \phi^{\rm PBCS}_h|\phi^{t\mbox{-}J}_h\rangle|^2 .
\end{align}
Let us call this the number-projected overlap squared (NPOS). 
As shown previously, the NPOS can provide a reasonable measure of the overlap except that there is a certain degree of arbitrariness in choosing the particle number.
In other words, the average particle number is already implicitly determined by the chemical potential $\mu$ in $H_{\rm PBCS}$.
Therefore, it is redundant to choose both particle number sector and chemical potential.

We fix this problem by weighting the NPOS over all possible particle number sectors:
\begin{align}
\overline{{\cal O}^2} = \sum_h |\lambda_h|^2 {\cal O}^2_h ,
\end{align}
which we call the number-weighted overlap squared (NWOS).
Note that the weighting factor can be computed via $|\lambda_h|^2 = \langle \psi_{\rm PBCS}| {\cal P}_h |\psi_{\rm PBCS} \rangle$, where ${\cal P}_h$ is the number projection operator.

It is instructive to perform an explicit analysis of the $2 \times 2$ system to demonstrate how the NWOS is actually implemented.
This analysis is also useful to illuminate the equivalence between the $t$-$J$ model and the projected BCS theory at half filling.
See {\bf Methods} for details.
Below, we present the numerical analysis of the $2 \times 2$ and larger systems via full-fledged exact diagonalization.

{\bf Unified phase diagram.}
Figure~\ref{fig:PBCSmu} shows the phase diagram of the $t$-$J$ model in the square lattice determined via the NWOS between the exact ground states of the $t$-$J$ model and the projected BCS theory with the $d$-wave pairing symmetry.
Note that the NWOS is computed as a function of $\Delta$ and $\mu$, both of which are varied for all possible values.
Also, note that $N_s$ is appropriately chosen for the proper tessellation of the square lattice.
See {\bf Methods} for details.

There are several important features to be noted.
First, as expected from the equivalence at half filling, the NWOS is unity regardless of $\Delta$ in the limit of large $\mu$.
This confirms that AF is precisely captured by the projected BCS theory.

Second, the NWOS becomes also unity in the limit of negatively large $\mu$, corresponding to the vacuum.
Here, the exact ground states of the $t$-$J$ model and the projected BCS theory are trivially identical since both spin exchange and pairing terms play negligible roles.
Near this limit, the ground state is the usual Fermi liquid state.

Finally, clearly being non-zero, the optimal $\Delta$ producing the maximum NWOS forms a rather well-defined curve as a function of $\mu$, especially for larger systems at $N_s=16$ and 20. 
Note that, at $N_s=18$, there is a large black dome around $\mu=0$, masking the otherwise high NWOS region. 
The high NWOS can be obtained in this region if the first excited state of the projected BCS theory, which is nearly degenerate with its exact ground state, is used to compute the NWOS. 
In conclusion, there is a well-defined curve of the non-zero pairing amplitude as a function of doping.

Actually, it is physically meaningful to express the phase diagram as a function of hole concentration $x$ rather than $\mu$.
To this end, we convert $\mu$ to $x$ by inverting the relationship $x = 1- \sum_i \langle n_i \rangle /N_s$ with $\langle n_i \rangle$ implicitly depending on $\mu$.
Figure~\ref{fig:PBCSx} shows the phase diagram of $t$-$J$ model in the square lattice as a function of $\Delta$ and $x$.
It is now possible to determine the optimal pairing amplitude producing the maximum NWOS, $\Delta_{\rm max}$, as a function of $x$.
As one can see from Fig.~\ref{fig:PBCSx}, $\Delta_{\rm max}$ forms a well-defined curve clearly lifted from $\Delta=0$, especially at $N_s=16$ and 20.

Note that the spin exchange coupling constant of the $t$-$J$ model is fixed to be $J/t=0.5$ in Figs.~\ref{fig:PBCSmu} and \ref{fig:PBCSx}.
See {\bf Methods} to find how the phase diagram changes with variation of $J/t$. 

Now, one may wonder if the non-zero value of $\Delta$ actually means superconductivity. 
To answer this question, we compute the superconducting order parameter by using the exact ground states of the projected BCS theory with $\Delta_{\rm max}$ determined above as a function of $x$.

{\bf Superconducting order parameter.}
One of the most advantageous features of the projected BCS theory is that its exact ground states have intrinsic particle number fluctuations. 
Thus, it is possible to directly compute the superconducting order parameter, $\langle c_{i\uparrow} c_{j\downarrow} \rangle$, instead of taking the large-distance limit of the off-diagonal long-range order (ODLRO), $\langle c_{i\uparrow} c_{j\downarrow} c^\dagger_{k\downarrow} c^\dagger_{l\uparrow} \rangle$, with $i$ $(k)$ and $j$ $(l)$ being the nearest neighbors, but $i$ and $k$ taken to be largely separated.

Specifically, we compute the superconducting order parameter as follows.
First, we obtain $\Delta_{\rm max}$ as a function of $x$ as shown in Fig.~\ref{fig:SC_order_parameter}~({\bf a}).
Then, we compute $\langle c_{i\uparrow} c_{j\downarrow} \rangle$ by using the exact ground states of the projected BCS theory with $\Delta_{\rm max}$ as an input parameter.
Figure~\ref{fig:SC_order_parameter}~({\bf b}) shows the resulting superconducting order parameter.
It is important to note that the superconducting order parameter vanishes as $x$ decreases in contrast to $\Delta_{\rm max}$, which approaches a finite value. 
This split between the pairing amplitude and the real superconducting order parameter is due to the strong correlation between Cooper pairs.

To elucidate the role of the strong correlation more clearly, we compare $\langle c_{i\uparrow} c_{j\downarrow} \rangle$ with the ``bare'' superconducting order parameter $\langle c_{i\uparrow} c_{j\downarrow} \rangle_0$, which would be the superconducting order parameter if there were no Gutzwiller projection:
\begin{align}
\langle c_{i\uparrow} c_{j\downarrow} \rangle_0 = \sum_{\bf k} \frac{\Delta_{0,{\bf k}} }{2E_{0,{\bf k}} } \cos{k_x},
\end{align}
where $E_{0,{\bf k}}=\sqrt{\xi^2_{0,{\bf k}}+\Delta^2_{0,{\bf k}} }$ with $\Delta_{0,{\bf k}}=2\Delta_{\rm max}(\cos{k_x}-\cos{k_y})$ and $\xi_{0,{\bf k}}=-2t(\cos{k_x}+\cos{k_y})-\mu_0$.
Above, the translational invariance is assumed.
Note that $\mu_0$ is determined from the condition that the ``bare'' hole concentration is the same as $x$: $\frac{1}{N_s}\sum_{\bf k} \xi_{0,{\bf k}}/E_{0,{\bf k}}=x$.
As shown in Fig.~\ref{fig:SC_order_parameter}~({\bf b}), the bare superconducting order parameter does not vanish at low doping similar to $\Delta_{\rm max}$.

While important to identify superconductivity, the superconducting order parameter is not a direct physical observable by itself. 
For a physical observable, the expectation value of the pairing energy, $\Delta_{\rm max} \langle c_{i\uparrow}c_{j\downarrow} \rangle$, is shown in Fig.~\ref{fig:SC_order_parameter}~({\bf c}).
Similar to the superconducting order parameter, the expectation value of the pairing energy vanishes at low doping while its bare counterpart, $\Delta_{\rm max} \langle c_{i\uparrow}c_{j\downarrow} \rangle_0$, does not.
It is interesting to mention that a similar split between the pairing amplitude and the real superconducting order parameter was previously observed~\cite{Paramekanti01, Paramekanti04}.

{\bf Comparison with the RVB state.}
To put the results of the projected BCS theory into prospective, we perform the similar NWOS analysis for the RVB state.
To this end, it is necessary to construct the RVB state precisely tailor-made for each finite system. 
Specifically, we would like to find the amplitude of the RVB state for each given basis state with spin up and down electrons located in $\{{{\bf r}_\uparrow}, {\bf r}_\downarrow\}$, respectively:
\begin{align}
{\cal A}_{\rm RVB}(\{{\bf r}_\uparrow,{\bf r}_\downarrow\})=\det{(\tilde{g}_{ij})}  ,
\label{eq:A_RVB}
\end{align}
where $\tilde{g}_{ij}=\tilde{g}({\bf r}_i - {\bf r}_j)$ is the Fourier transform of $g_{\bf k}=\Delta_{\bf k}/(\xi_{\bf k}+E_{\bf k})$ 
with $E_{\bf k}=\sqrt{\xi^2_{\bf k}+\Delta^2_{\bf k}}$, $\Delta_{\bf k}=2\Delta(\cos{k_x}-\cos{k_y})+\delta$, and $\xi_{\bf k}=-2t(\cos{k_x}+\cos{k_y})-\mu$.
The indices $i$ and $j$ run though all the coordinates of spin up and down electrons in $\{{{\bf r}_\uparrow}, {\bf r}_\downarrow\}$, respectively. 
Note that a small constant $\delta$ is added to the pairing term $\Delta_{\bf k}$ for technical convenience.
See {\bf Methods} for the detailed derivation of Eq.~\eqref{eq:A_RVB}.

Figure~\ref{fig:RVBmu} shows the phase diagram of the $t$-$J$ model in the square lattice determined via the NWOS between the exact ground state of the $t$-$J$ model and the RVB state with the $d$-wave pairing symmetry.
Due to its inability to capture the long-range antiferromagnetic order, the RVB state has zero NWOS with the exact ground state of the $t$-$J$ model at sufficiently large $\mu$ for $N_s=4, 10, 16, 20$.
For $N_s=8$ and 18, the RVB state is not even available beyond certain large $\mu$, where the Gutzwiller projection completely annihilate the BCS state, making the RVB state a null state.

Similar to what is done for the projected BCS theory, the RVB phase diagram can be also plotted as a function of $x$ in Fig.~\ref{fig:RVBx}.
Again, the RVB state has zero NWOS with the exact ground state of the $t$-$J$ model at half filling.
Away from half filling, overall, the RVB phase diagram shows irregular behaviors depending on $N_s$ with broad vacant regions.
Even when the RVB phase diagram behaves reasonably well at $N_s=10$ and 20, the optimal pairing amplitude of the RVB state exhibits a rather different behavior from that of the projected BCS theory at low doping.
That is, the optimal pairing amplitude of the RVB state vanishes as $x$ approaches zero.

Before discussing the different behaviors of the optimal pairing amplitude, however, it is important to check which among the two states, the RVB state and the exact ground state of the projected BCS theory, has the higher overlap with the exact ground state of the $t$-$J$ model as a function of $x$.
Figure~\ref{fig:NWOS_max} shows the maximum NWOS, $\overline{{\cal O}^2}_{\rm max}$, of the projected BCS theory in comparison with that of the RVB state.
As one can see, $\overline{{\cal O}^2}_{\rm max}$ of the projected BCS theory is pinned to be exactly unity at half filling, while that of the RVB state approaches a seemingly random value as $x \rightarrow 0$ and becomes entirely zero strictly at $x=0$.
More importantly, $\overline{{\cal O}^2}_{\rm max}$ of the projected BCS theory is reduced somewhat upon initial doping but bounces back to a high value as $x$ increases. 
Meanwhile, $\overline{{\cal O}^2}_{\rm max}$ of the RVB state is generally lower than that for the projected BCS theory except for a narrow range of $x$ at low doping.  
It is interesting to speculate if the RVB state can provide a good trial state at this range of doping, where the spin glass phase can occur.

In summary, the projected BCS theory provides better trial states for the ground state of the $t$-$J$ model than the RVB state, capturing both AF at half filling and SCSC at moderate doping.

{\bf S-wave pairing symmetry.}
As mentioned previously, the equivalence at half filling does not necessarily guarantee a high overlap upon doping, while so for the $d$-wave pairing symmetry in the square lattice.
We would like to investigate what happens for the $s$-wave pairing symmetry in the square lattice. 

Figure~\ref{fig:s-wave} shows that, in this case, the NWOS is always maximized along $\Delta=0$ for all finite $\mu$, meaning that the $s$-wave pairing cannot be formed at finite doping.
Note that the NWOS is unity regardless of $\Delta$ in both limits of $\mu=\infty$ and $-\infty$, corresponding to half filling and vacuum, respectively.
According to the proof of equivalence, the projected BCS theory is entirely equivalent to the Heisenberg model at half filling regardless of the pairing symmetry.
The vacuum states are trivially identical.

It is important to note that the maximum NWOS of the $s$-wave pairing symmetry along $\Delta=0$ is bound to be lower than that of the $d$-wave pairing symmetry occurring at the optimal curve of $\Delta$ in Fig.~\ref{fig:PBCSmu}, considering both pairing symmetries generate the same Hamiltonian at $\Delta=0$.
In conclusion, the $d$-wave pairing symmetry is preferred over the $s$-wave counterpart for the $t$-$J$ model in the square lattice.

{\bf Geometrical frustration.}
Now, we investigate how the overlap depends on the lattice structure, especially with geometrical frustration.
Specifically, we perform the NWOS analysis for the projected BCS theory in the triangular lattice. 

Similar to the square lattice, we consider both $s$- or $d$-wave pairing symmetries.
While simply a constant for the $s$-wave pairing, the pairing amplitude is given as $\Delta_{ij}=\Delta e^{2i\theta_{ij}}$ for the $d$-wave pairing with $\theta_{ij}$ being the angle between the $x$ axis and the nearest-neighbor-connecting vector ${\bf r}_{ij}={\bf r}_i-{\bf r}_j$.
Figure~\ref{fig:Triangular} shows that, at moderate doping, the projected BCS theory provides poor trial states for the $t$-$J$ model in the triangular lattice regardless of the pairing symmetry. 

In conclusion, geometrical frustration, at least, in the form of the triangular lattice is detrimental to the formation of superconductivity in contrast to the original rationale behind the RVB state.
This result is consistent with the experimental observation that high-temperature superconductivity occurs almost always in tandem with unfrustrated AF at half filling.

{\bf Discussion}

In this work, it is shown that the projected BCS theory with the $d$-wave pairing symmetry can provide excellent trial states for the exact ground states of the $t$-$J$ model in the square lattice for a wide range of hole concentration.
In particular, the unified phase diagram, capturing both AF at half filling and SCSC at moderate doping, is obtained by computing the overlap between the exact ground states of the $t$-$J$ model and the projected BCS theory with the $d$-wave pairing symmetry.
A main breakthrough in this work is to devise the overlap properly taking into account particle number fluctuations. 

By utilizing the natural presence of particle number fluctuations, the superconducting order parameter can be computed directly from the exact ground state of the projected BCS theory.
The split between the optimal pairing amplitude and the superconducting order parameter can in principle provide an explanation for the pseudogap phenomenon at low doping.

For future work, it would be interesting to investigate various spectral properties of the projected BCS theory by computing both normal and anomalous Green's functions in the presence of intrinsic particle number fluctuations. 
This investigation could shed light on the issue of strange metal behaviors.

{\bf Methods}

{\bf Symmetries.}
Group theoretical analyses can facilitate exact diagonalization. 
Specifically, the basis states expanding the Hilbert space can be conveniently categorized into appropriate symmetry sectors, block-diagonalizing the Hamiltonian.

In this work, we are interested in the sub-Hilbert space with both total momentum and $z$-component of the total spin being zero.
In this situation, it is sufficient to focus on the point group symmetries.

Our goal is to understand the point group symmetries of the $t$-$J$ and projected BCS Hamiltonians.
Note that, being a local operator, the Gutzwiller projection commutes with all the point group symmetries. 
Therefore, one does not need to consider the Gutzwiller projection for the discussion of the point group symmetries.

Let us first discuss the point group symmetries in the square lattice.
The point groups representing the symmetries of the square lattice are given by $\mathcal{C}_{4{\rm v}} = \{ e, C_4, C_2, C_4^3, \sigma_x, \sigma_y, \sigma_{\rm d}^{+}, \sigma_{\rm d}^{-} \}$, where $e$ is the identity operator, $C_n$ the rotation operator about the $z$ axis by the angle of $2\pi / n$, $\sigma_{x/y}$ the reflection operator with respect to the $x/y$ axis, and $\sigma_{\rm d}^{+/-}$ the reflection operator with respect to the $y = \pm x$ line.
With spin degrees of freedom, the point groups should be enlarged to include the magnetic (or color) group~\cite{LudwigFalter}.
In other words, one needs to consider the effects of the spin-flip operator $f$.

Specifically, the point groups of the $t$-$J$ model in the square lattice can be determined by the fact that both $H_t$ and $H_J$ commute with all the symmetry operators in $\mathcal{C}_{4{\rm v}}$ as well as the spin-flip operator $f$.
In other words, the $t$-$J$ model has $\mathcal{C}_{4{\rm v}}^{\rm II} \equiv  \{e, f\} \bigotimes  \mathcal{C}_{4{\rm v}}$ as its point groups in the square lattice.

Meanwhile, the point groups of the projected BCS Hamiltonian depend on the pairing symmetry.
To understand this, it is important to note that $H_\Delta$ does not commute with some of the symmetry operators in $\mathcal{C}_{4{\rm v}}^{\rm II}$, while both $H_t$ and $H_\mu$ do so with all of them.
Specifically, $f H_\Delta f^{-1} = -H_\Delta$ for both $s$- and $d$-wave pairing symmetries.
Also, $C_4 H_\Delta C_4^{-1} = -H_\Delta^{}$ for the $d$-wave pairing symmetry in the square lattice.

Consequently, in the square lattice, the point groups of the projected BCS Hamiltonian with the $s$-wave pairing symmetry are simply given by $\mathcal{C}_{4{\rm v}}^{\rm I} \equiv \mathcal{C}_{4{\rm v}}$, excluding the spin-flip operator.
In the case of the $d$-wave pairing symmetry, the spin-flip operator can be combined with some of the symmetry operators in $\mathcal{C}_{4{\rm v}}$.
As a result, it can be shown that the point groups of the projected BCS Hamiltonian with the $d$-wave pairing symmetry are given by $\mathcal{C}_{4{\rm v}}^{\rm III} \equiv \{e, C_2, \sigma_x, \sigma_y, fC_4, fC_4^3, f\sigma_{\rm d}^{+}, f\sigma_{\rm d}^{-}\}$ in the square lattice.

Note that the point groups of the projected BCS Hamiltonian ($\mathcal{C}_{4{\rm v}}^{\rm I}$ and $\mathcal{C}_{4{\rm v}}^{\rm III}$ for the $s$- and $d$-wave pairing symmetries, respectively) are only the subgroups of those of the $t$-$J$ Hamiltonian ($\mathcal{C}_{4{\rm v}}^{\rm II}$).
In general, if two Hamiltonians have different symmetries, the overlap between their ground states would be suppressed.

Fortunately, the effects of both $f$ and $C_4$ on $H_\Delta$ can be absorbed into the sign change of the pairing amplitude $\Delta$. 
The sign, more generally, the phase of the pairing amplitude only affects the relative phase of the superconducting ground state between different particle number sectors, while keeping the component of the state in each sector invariant.
Consequently, if properly defined to take into account particle number fluctuations, the overlap would not be automatically suppressed.  
See the main text for details.

Finally, let us consider the point group symmetries in the triangular lattice.
The point groups representing the symmetries of the triangular lattice are $\mathcal{C}_{6{\rm v}}$.
Since $f$ still works as a symmetry element for the $t$-$J$ model, the point groups of the $t$-$J $model are given by $\mathcal{C}_{6{\rm v}}^{\rm II} \equiv \{e, f\} \bigotimes  \mathcal{C}_{6{\rm v}}$.
Similar to the square lattice, the projected BCS Hamiltonian with the $s$-wave pairing symmetry has $\mathcal{C}_{6{\rm v}}^{\rm I} \equiv \mathcal{C}_{6{\rm v}}$ in the triangular lattice due to the lack of the spin-flip symmetry.
Even worse, the projected BCS Hamiltonian with the $d$-wave pairing symmetry has only $\mathcal{C}_{2}$ as its point groups in the triangular lattice.

{\bf Proof of equivalence at half filling.}
The energy eigenstates of the projected BCS theory can be expanded in terms of their component in each particle number sector:
\begin{align}
|\psi\rangle = \sum_{h=0}^{N_s} \lambda_h |\phi_h\rangle,
\label{eq:hole_expansion}
\end{align}
where $h$ is the hole number, related with the site number $N_s$ and the electron number $N_e$ via $h=N_s-N_e$. 
$\lambda_h$ is the weighting amplitude in each hole number sector. 
Note that $h$ is always an even number since we are interested in the paired state. 

For convenience, let us rewrite the Schr\"{o}dinger equation for the projected BCS theory as follows: 
\begin{align}
H_{\rm PBCS} = {\cal H}_t +{\cal H}_c +{\cal H}_a -\mu N,
\end{align}
where ${\cal H}_{t,c,a}={\cal P}_G H_{t,c,a} {\cal P}_G$ with $H_c$ and $H_a$ being the creation and annihilation parts of $H_\Delta$, respectively.
The particle number operator is defined as $N=\sum_i n_i$.

Then, the Schr\"{o}dinger equation can be written component by component:
(i) for $h=0$,
\begin{align}
\lambda_2 {\cal H}_c|\phi_2\rangle =\lambda_0 (E+N_s\mu)|\phi_0\rangle ,
\label{eq:h_0}
\end{align}
(ii) for $2 \leq h \leq N_s-2$,
\begin{align}
\lambda_h {\cal H}_t |\phi_h\rangle &+\lambda_{h+2}{\cal H}_c|\phi_{h+2}\rangle +\lambda_{h-2}{\cal H}_a|\phi_{h-2}\rangle \nonumber \\
&=\lambda_h (E+\mu(N_s-h))|\phi_h\rangle ,
\label{eq:2_leq_h_leq_Ns-2}
\end{align}
and (iii) for $h=N_s$,
\begin{align}
\lambda_{N_s-2}{\cal H}_a|\phi_{N_s-2}\rangle = \lambda_{N_s}E|\phi_{N_s}\rangle .
\label{eq:h_Ns-2}
\end{align}

Half filling can be obtained in the limit of infinite chemical potential, $\mu \rightarrow \infty$.
In this limit, to satisfy Eq.~\eqref{eq:h_0}, $E=-N_s\mu +{\cal O}(1/\mu^\alpha)$ with $\alpha \geq 0$.
Note that $\lambda_h$ cannot diverge since they are the coefficients of a normalized wave function. 
Meanwhile, according to Eq.~\eqref{eq:h_Ns-2}, ${\cal O}({\lambda_{N_s-2}})= {\cal O}(\mu{\lambda}_{N_s})$.
Under the assumption that $\lambda_h={\cal O}(1/\mu^{\nu_h})$, this means that $\nu_{N_s}=\nu_{N_s-2}+1$.
Similarly, according to Eq.~\eqref{eq:2_leq_h_leq_Ns-2}, $\nu_h=\nu_{h-2}+1$ for $h \geq 2$.
Considering that $\nu_0=0$ (i.e., $h_0={\cal O}(1)$) due to the wave function normalization, it can be concluded that $\nu_h=h/2$.
This, in turn, means that one can now determine $\alpha$ in the scaling of $E$: $\alpha=1$.
Therefore, $E=-N_s\mu +\xi/\mu +{\cal O}(1/\mu^\beta)$ with $\beta >1$.

In this scaling, one can rewrite the two component-by-component equations for $h=0$ and $2$ by keeping only the most dominant terms:
\begin{align}
\lambda_2 {\cal H}_c |\phi_2\rangle &= \lambda_0 \frac{\xi}{\mu}|\phi_0\rangle, \\
\lambda_0 {\cal H}_a |\phi_0\rangle &= -2\lambda_2\mu |\phi_2\rangle,
\end{align}
which can be combined to generate the Schr\"{o}dinger equation solely at $h=0$ as follows:
\begin{align}
{\cal H}_c {\cal H}_a |\phi_0\rangle = -2\xi|\phi_0\rangle .
\end{align}

Under the no-double occupancy constraint imposed by the Gutzwiller projection, one can show that 
\begin{align}
{\cal H}_c {\cal H}_a |\phi_0\rangle &= \Delta^2 \sum_{\langle i,j \rangle} (c^\dagger_{i\uparrow} c^\dagger_{j\downarrow} -c^\dagger_{i\downarrow} c^\dagger_{j\uparrow})
(c_{j\downarrow} c_{i\uparrow} -c_{j\uparrow} c_{i\downarrow})  |\phi_0\rangle \nonumber \\
&= -2\Delta^2 \sum_{\langle i,j \rangle} \left( {\bf S}_i \cdot {\bf S}_j -n_i n_j /4 \right) |\phi_0\rangle,
\end{align}
where the second line is obtained by examining the matrix elements for all possible spin configurations in the $(i,j)$ sites, i.e., $|\uparrow\uparrow\;\rangle, |\uparrow\downarrow\;\rangle, |\downarrow\uparrow\;\rangle, |\downarrow\downarrow\;\rangle$.
Consequently, the final Schr\"{o}dinger equation at half filling becomes as follows:
\begin{align}
\Delta^2 \sum_{\langle i,j \rangle} \left( {\bf S}_i \cdot {\bf S}_j -n_i n_j /4 \right) |\phi_0\rangle = \xi|\phi_0\rangle ,
\end{align}
which is nothing but the Hamiltonian of the Heisenberg model.

It is interesting to note that the equivalence between the projected BCS theory and the Heisenberg model at half filling has been previously argued by using a different analytical method~\cite{Park05_PRB}, where the no-double-occupancy constraint is relaxed so that the projected BCS Hamiltonian is replaced by the BCS Hamiltonian with finite on-site repulsive interaction $U$. 
Eventually, the equivalence is obtained in the limit of large $U$.
Note that the equivalence has been also argued by using a path-integral approach~\cite{Kochetov09}.

{\bf Explicit analysis of the $2 \times 2$ system.}
The entire Hilbert space of the $2 \times 2$ system can be expanded by the following seven basis states:
(i) three states in the half-filled sector,
\begin{align}
|e_1\rangle &= \frac{1}{\sqrt{2}} \left( \Big|
\begin{array}{cc}
\uparrow  &  \downarrow \\
\downarrow  &  \uparrow  
\end{array}
\Big\rangle
+\Big|
\begin{array}{cc}
\downarrow  &  \uparrow \\
\uparrow  &  \downarrow  
\end{array}
\Big\rangle \right), 
\nonumber \\
|e_2\rangle &= \frac{1}{2} \left( \Big|
\begin{array}{cc}
\uparrow  &  \uparrow \\
\downarrow  &  \downarrow  
\end{array}
\Big\rangle
+\Big|
\begin{array}{cc}
\downarrow  &  \downarrow \\
\uparrow  &  \uparrow  
\end{array}
\Big\rangle 
+\Big|
\begin{array}{cc}
\downarrow  &  \uparrow \\
\downarrow  &  \uparrow  
\end{array}
\Big\rangle
+\Big|
\begin{array}{cc}
\uparrow  &  \downarrow \\
\uparrow  &  \downarrow  
\end{array}
\Big\rangle 
\right), 
\nonumber \\ 
|e_3\rangle &= \frac{1}{2} \left( \Big|
\begin{array}{cc}
\uparrow  &  \uparrow \\
\downarrow  &  \downarrow  
\end{array}
\Big\rangle
+\Big|
\begin{array}{cc}
\downarrow  &  \downarrow \\
\uparrow  &  \uparrow  
\end{array}
\Big\rangle 
-\Big|
\begin{array}{cc}
\downarrow  &  \uparrow \\
\downarrow  &  \uparrow  
\end{array}
\Big\rangle
-\Big|
\begin{array}{cc}
\uparrow  &  \downarrow \\
\uparrow  &  \downarrow  
\end{array}
\Big\rangle 
\right),
\end{align}
(ii) three states in the two-hole sector,
\begin{align}
|e_4\rangle &= \frac{1}{2} \Bigg( \Big|
\begin{array}{cc}
0  &  \uparrow \\
\downarrow  &  0
\end{array}
\Big\rangle
+\Big|
\begin{array}{cc}
\uparrow  &  0 \\
0  &  \downarrow  
\end{array}
\Big\rangle 
-\Big|
\begin{array}{cc}
\downarrow  &  0 \\
0  &  \uparrow  
\end{array}
\Big\rangle
-\Big|
\begin{array}{cc}
0 &  \downarrow \\
\uparrow  &  0  
\end{array}
\Big\rangle 
\Bigg), 
\nonumber \\ 
|e_5\rangle &= \frac{1}{\sqrt{8}} \Bigg( \Big|
\begin{array}{cc}
0  &  0 \\
\downarrow  &  \uparrow  
\end{array}
\Big\rangle
-\Big|
\begin{array}{cc}
0  &  0 \\
\uparrow  &  \downarrow  
\end{array}
\Big\rangle 
+\Big|
\begin{array}{cc}
\downarrow  &  \uparrow \\
0  &  0  
\end{array}
\Big\rangle
-\Big|
\begin{array}{cc}
\uparrow  &  \downarrow \\
0  &  0  
\end{array}\Big\rangle 
\nonumber \\
&+\Big|
\begin{array}{cc}
\uparrow  &  0 \\
\downarrow  &  0  
\end{array}
\Big\rangle
-\Big|
\begin{array}{cc}
\downarrow  &  0 \\
\uparrow  &  0  
\end{array} 
\Big\rangle 
+\Big|
\begin{array}{cc}
0  &  \uparrow \\
0  &  \downarrow  
\end{array}
\Big\rangle
-\Big|
\begin{array}{cc}
0  &  \downarrow \\
0  &  \uparrow  
\end{array}
\Big\rangle\Bigg),
\nonumber \\
|e_6\rangle &= \frac{1}{\sqrt{8}} \Bigg( \Big|
\begin{array}{cc}
0  &  0 \\
\downarrow  &  \uparrow  
\end{array}
\Big\rangle
-\Big|
\begin{array}{cc}
0  &  0 \\
\uparrow  &  \downarrow  
\end{array}
\Big\rangle 
+\Big|
\begin{array}{cc}
\downarrow  &  \uparrow \\
0  &  0  
\end{array}
\Big\rangle
-\Big|
\begin{array}{cc}
\uparrow  &  \downarrow \\
0  &  0  
\end{array}\Big\rangle 
\nonumber \\
&-\Big|
\begin{array}{cc}
\uparrow  &  0 \\
\downarrow  &  0  
\end{array}
\Big\rangle
+\Big|
\begin{array}{cc}
\downarrow  &  0 \\
\uparrow  &  0  
\end{array} 
\Big\rangle 
-\Big|
\begin{array}{cc}
0  &  \uparrow \\
0  &  \downarrow  
\end{array}
\Big\rangle
+\Big|
\begin{array}{cc}
0  &  \downarrow \\
0  &  \uparrow  
\end{array}
\Big\rangle\Bigg), 
\end{align}
and finally (iii) one state in the vacuum sector,
\begin{align}
|e_7\rangle &= \Big|
\begin{array}{cc}
0  &  0 \\
0  &  0  
\end{array}
\Big\rangle.
\end{align}
Note that we are interested in the Hilbert space with zero momentum (i.e., translationally invariant) and zero $z$-component of the total spin (i.e., spin-flip invariant).

In terms of the ordered basis set $\{ |e_1\rangle, |e_2\rangle, |e_3\rangle \}$, the $t$-$J$ Hamiltonian can be written in the half-filled sector as follows:
\begin{align}
H_{t\mbox{-}J} = 2J
\left(
\begin{array}{ccc}
-2          &  \sqrt{2}   &        0   \\
\sqrt{2}  &  -1           &        0   \\
0           &	0             &	-1
\end{array}
\right),
\end{align}
generating the following ground state:
\begin{align}
|\phi^{t\mbox{-}J}_0\rangle = {\cal N}_0 \left( |e_1\rangle -\frac{1}{\sqrt{2}}|e_2\rangle \right),
\label{eq:phi_t-J_0}
\end{align}
where ${\cal N}_0=\sqrt{2/3}$ is the normalization constant.
Note that $|e_3\rangle$ couples with neither $|e_1\rangle$ nor $|e_2\rangle$ since they have different point group symmetries.

Similarly, in terms of the ordered basis set $\{ |e_4\rangle, |e_5\rangle, |e_6\rangle \}$, the $t$-$J$ Hamiltonian can be written in the two-hole sector as follows:
\begin{align}
H_{t\mbox{-}J} = -2
\left(
\begin{array}{ccc}
0	         & 2\sqrt{2}t	& 0   \\
2\sqrt{2}t	& J	                 & 0   \\
0	         & 0	                 & J
\end{array}
\right),
\end{align}
generating the following ground state:
\begin{align}
|\phi^{t\mbox{-}J}_2\rangle = {\cal N}_2^{} \Big(
\alpha |e_4\rangle + \beta |e_5\rangle 
\Big) ,
\label{eq:phi_t-J_2}
\end{align}
where $\alpha=\sqrt{J^2 + 32 t^2} - J$, $\beta= 4\sqrt{2} t$, and ${\cal N}_2=1/\sqrt{|\alpha|^2+|\beta|^2}$. 
Similar to half filling, $|e_6\rangle$ couples with neither $|e_4\rangle$ nor $|e_5\rangle$.

Meanwhile, the projected BCS Hamiltonian with the $d$-wave pairing symmetry can be written as the following block diagonal form in terms of the ordered basis set $\{ |e_1\rangle, |e_2\rangle, |e_4\rangle, |e_5\rangle, |e_3\rangle, |e_6\rangle, |e_7\rangle\}$:
\begin{align}
H_{\rm PBCS} &= 
\left(
\begin{array}{cccc}
-4\mu    &  0                        &  0                  & 4\Delta                 \\
0           &  -4\mu                 &  0                  & -2\sqrt{2}\Delta     \\
0           &  -0                       &  -2\mu           & -4\sqrt{2}t             \\
4\Delta  &  -2\sqrt{2}\Delta  &  -4\sqrt{2}t    & -2\mu                      
\end{array}
\right) 
\nonumber \\
&\oplus \left(
\begin{array}{ccc}
-4\mu & 2\sqrt{2}\Delta & 0\\
2\sqrt{2}\Delta & -2\mu & -4\sqrt{2}\Delta\\
0 & -4\sqrt{2}\Delta & 0
\end{array}
\right).
\end{align} 
The characteristic polynomial of each block matrix is given as follows:
\begin{align}
\mathcal{C}_{-}(\varepsilon) &= (\varepsilon + 4\mu) \Big[ (\varepsilon + 4\mu)^3 - 4\mu (\varepsilon + 4\mu)^2 
\nonumber \\
&+4 (\mu^2 - 6 \Delta^2 - 8 t^2) (\varepsilon + 4\mu) + 48 \Delta^2 \mu \Big], 
\nonumber \\
\mathcal{C}_{+}(\varepsilon) &= (\varepsilon + 4\mu)^3 - 6 \mu (\varepsilon + 4\mu)^2 
\nonumber \\
&+ 8 (\mu^2 - 5 \Delta^2) (\varepsilon + 4\mu) + 32\Delta^2\mu,
\end{align}
where the subscript $-/+$ indicates the parity associated with the $\pi/2$ rotation followed by the spin flip.

After a careful examination, one can show that the ground state occurs in the negative parity sector if $\mu$ gets sufficiently large.
In this situation, it is convenient to observe that there is an easy solution of $\mathcal{C}_{-}(\varepsilon)$, whose eigenvalue is $-4\mu$ with the corresponding eigenstate given by $|e_1\rangle + \sqrt{2} |e_2\rangle$. 
The other three eigenstates in the negative parity sector including the ground state should be orthogonal to this eigenstate.
Therefore, the ground state at sufficiently large $\mu$ takes the following form:
\begin{align}
|\psi_{\rm PBCS}\rangle = {\cal N} \left(
|e_1\rangle -\frac{1}{\sqrt{2}} |e_2\rangle + f |e_4\rangle  +  g |e_5\rangle
\right),
\label{eq:psi_PBCS}
\end{align}
where ${\cal N}=1/\sqrt{3/2+|f|^2+|g|^2}$ is the normalization constant, and $f$ and $g$ are some appropriate functions of $\Delta / t$ and $\mu / t$, whose details are not provided here except that $|f|, |g| \ll 1$ for sufficiently large $\mu$.
It is important to note that $|\psi_{\rm PBCS}\rangle$ has the component of $|e_1\rangle -\frac{1}{\sqrt{2}} |e_2\rangle$ in the half-filled sector, which coincides exactly with the ground state of $t$-$J$ model in the same sector, $|\phi_0^{t\mbox{-}J}\rangle$, as expected from the proof of equivalence at half filling.

Actually, it is interesting to perform the similar analysis for the projected BCS theory with the $s$-wave pairing symmetry, whose ground state also coincides exactly with the ground state of $t$-$J$ model at half filling for basically the same reason mentioned above.
Thus, there is a certain degree of robustness in the equivalence between the projected BCS theory and the Heisenberg model at half filling. 

Now, an important question is if $|\psi_{\rm PBCS}\rangle$ can capture the ground state of $t$-$J$ model accurately as $\mu$ is reduced. 
That is, how accurately $f$ and $g$ in Eq.~\eqref{eq:psi_PBCS} can capture $\alpha$ and $\beta$ in Eq.~\eqref{eq:phi_t-J_2} as a function of $\mu$?
This can be checked via the NWOS:
\begin{align}
\overline{{\cal O}^2} = {\cal N}^2 \left(
\frac{3}{2} +\frac{|f^*\alpha+g^*\beta|^2}{|\alpha|^2+|\beta|^2} 
\right) .
\end{align}
As one can see from Fig.~\ref{fig:PBCSmu} at $N_s=4$, the NWOS remains high upon doping. 
Note that there is a phase transition along the certain critical line of $\mu$ and $\Delta$, beyond which the ground state of the projected BCS theory is no longer given by Eq.~\eqref{eq:psi_PBCS}.

While providing various successful small-system checks, the $2 \times 2$ system is way too small to determine if the optimal pairing amplitude can become non-zero at finite doping.
To this end, one has to analyze larger systems by using full-fledged exact diagonalization.

{\bf Tessellation of the lattice.}
Let us begin with the tessellation of the square lattice, which can be tiled with tilted square unit cells.
The tilted square unit cells are considered to observe the rotational symmetry of the lattice.

The distance between any two adjacent tilted square unit cells can be written as $d_{\rm sq}=\sqrt{n^2 + m^2}$ (in units of lattice constant) for any non-negative integers $n$ and $m$ without the loss of generality, which is also the side length of the titled square unit cell.  
This means that the number of sites included in the tilted square unit cell is $N_s = d_{\rm sq}^2= n^2 + m^2$.
Since we are interested in the systems with even number of particles with zero $z$-component of the total spin, both $m$ and $n$ should be either even or odd integers; $N_s=2, 4, 8, 10, 16, 18, 20, 26, 32, 34, 36, \cdots$.

\begin{figure}
\centering
\includegraphics[width=0.8\columnwidth]{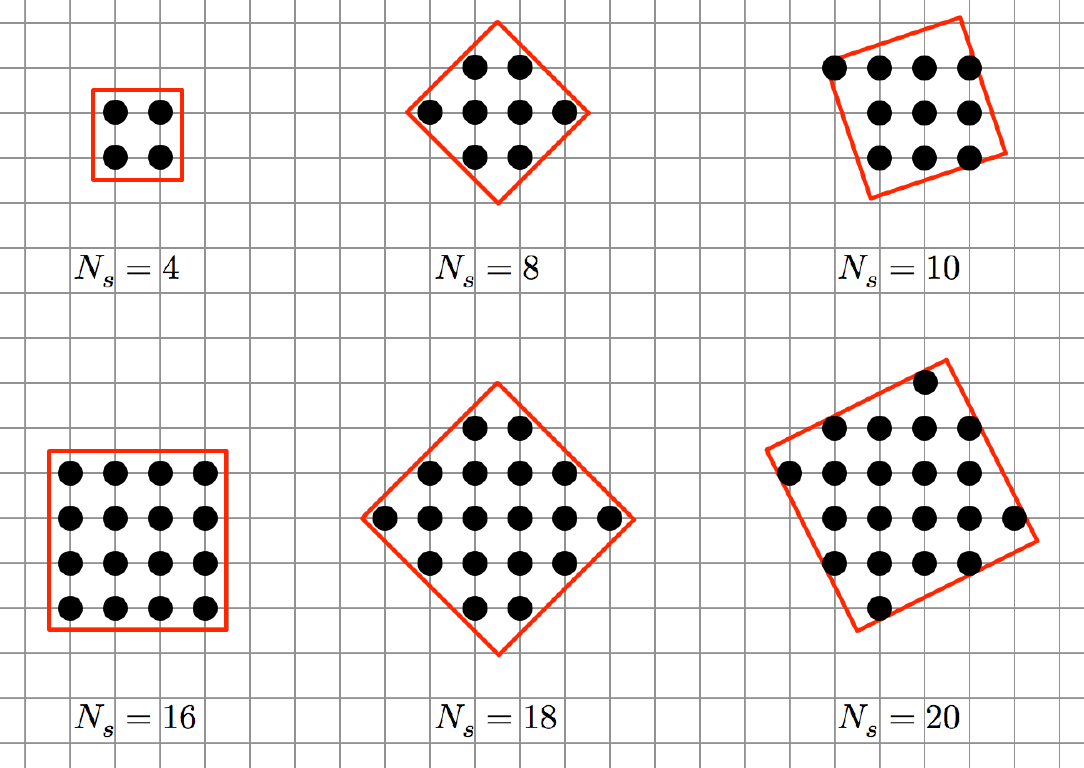}
\caption{{\bf Tessellation of the square lattice via tilted square unit cells.}
The point groups for $N_s =4, 8,16,18$ and $10, 20$ are $\mathcal{C}_{4{\rm v}}$ and $\mathcal{C}_4$, respectively.}
\label{fig:Tilted_square_tiling}
\end{figure}

For illustration, some of the tilted square unit cells are depicted in Fig.~\ref{fig:Tilted_square_tiling}.
Note that, if either $n$ or $m$ is zero (e.g., $N_s=4$ and 16), or $n=m$ (e.g., $N_s=8$ and 18), the tessellation with tilted square unit cells has additional reflection symmetries; one for the horizontal axis, one for the vertical axis, and two for the diagonal axes.
That is, the point groups become $\mathcal{C}_{4{\rm v}}$ for $N_s=4, 8, 16, 18$.

\begin{figure}
\centering
\includegraphics[width=0.8\columnwidth]{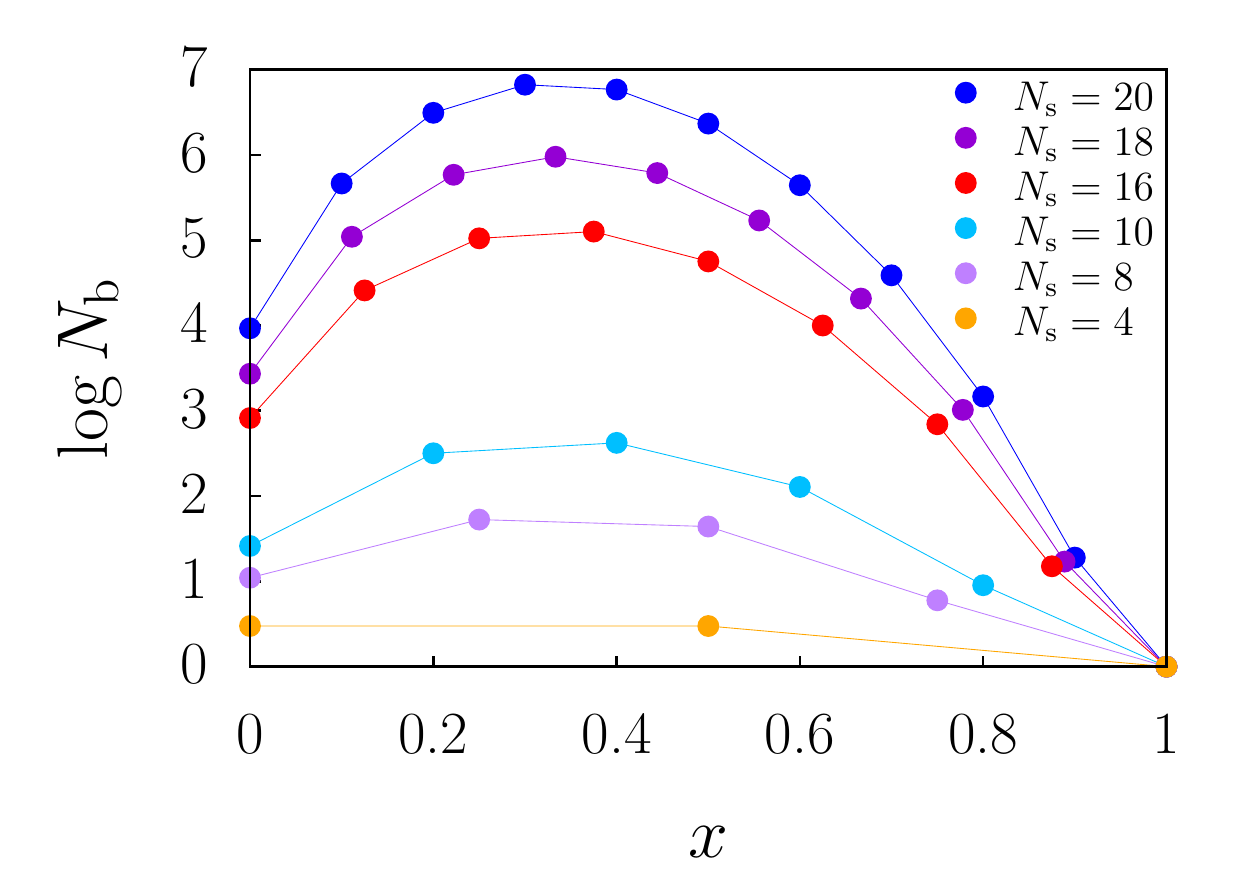}
\caption{{\bf Number of basis states for various tilted square unit cells as a function of hole concentration.}
Note that the ordinate is the number of basis states in the common logarithm scale, $\log{N_b}$.
Here, the hole concentration is defined only discretely as $x=1-N_e/N_s$ with $N_e$ and $N_s$ being the numbers of electrons and sites, respectively.
}
\label{fig:No_of_basis_states}
\end{figure}

Due to the exponential growth of the Hilbert space, we are able to perform exact diagonalization only up to $N_s=20$ in this work.
Figure~\ref{fig:No_of_basis_states} shows the number of basis states in the common logarithm scale, $\log{N_b}$, as a function of hole concentration $x=1-N_e/N_s$.
Note that, here, the number of basis states is computed within the restricted Hilbert space, where both total momentum and $z$-component of the total spin are zero without using any other point group symmetries.
As one can see, the total number of basis states can be more than 10 millions for $N_s=20$.
Roughly speaking, the total number of states increases by one order of magnitude as the number of sites increases by two.
This means that the next available system at $N_s=26$ would have more than 10 billion basis states, which are beyond the current computing capacity.

Now, let us switch gears to the tessellation of the triangular lattice.
Similar to the square lattice, the distance between any two adjacent tilted hexagonal unit cells can be written as $d_{\rm tr}=\sqrt{n^2 + nm + m^2}$ (in units of lattice constant) for any non-negative integers $n$ and $m$ without the loss of generality. 
After a moment of deliberation, one can show that the number of sites included in the tilted hexagonal unit cell is also given by $N_s =d_{\rm tr}^2 =n^2 + nm + m^2$.
Both $m$ and $n$ should be even integers for $N_s$ to be an integer number; $N_s = 4, 12, 16, 28, 36, \cdots$.
Meanwhile, the 120-degree spin order can exist only when $m = n\ ({\rm mod}\ 3)$. 
Combined all together, this means that the only accessible system in this work via exact diagonalization is that with $N_s=12$.

\begin{figure}
\centering
\includegraphics[width=0.7\columnwidth]{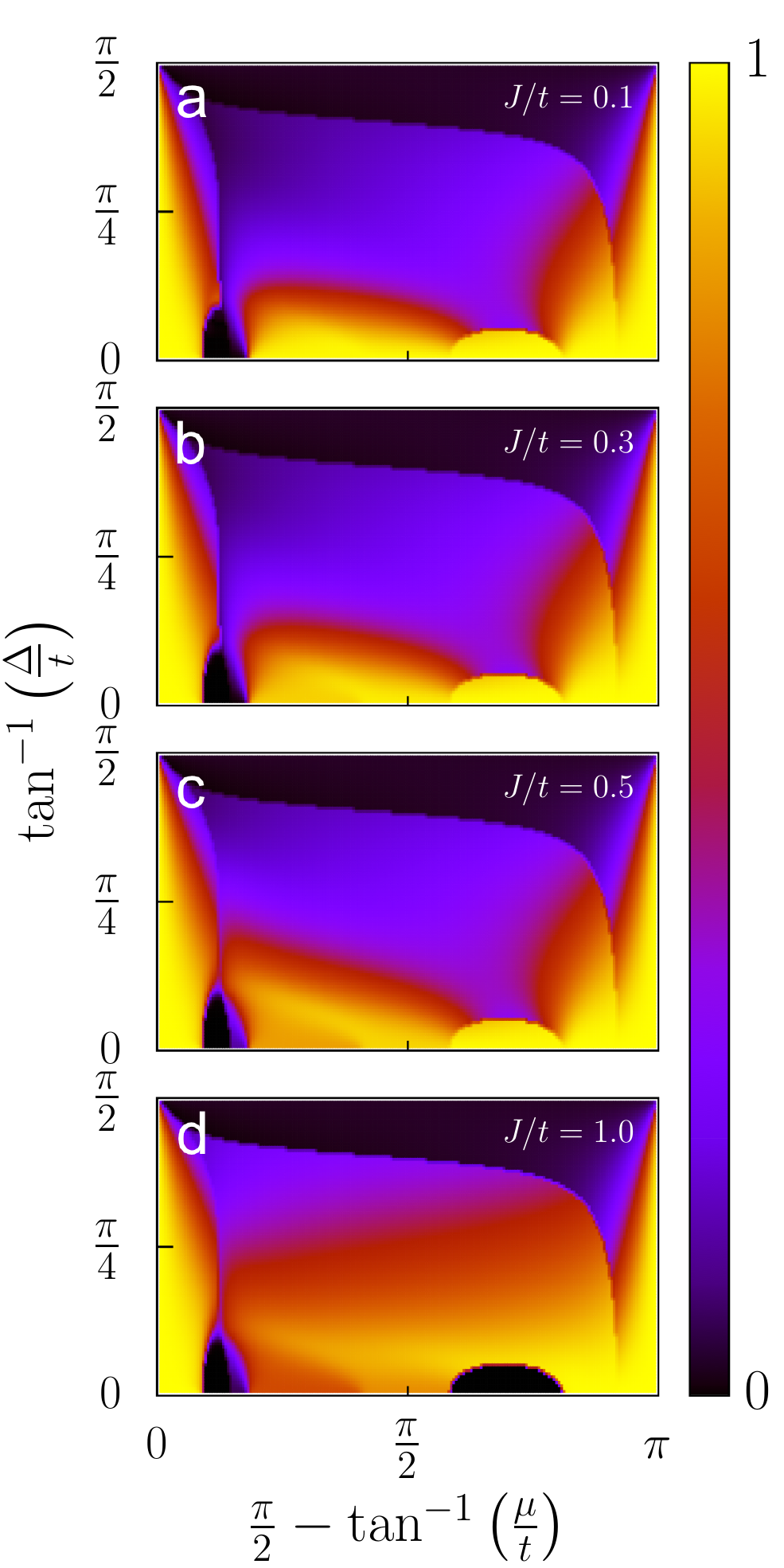}
\caption{{\bf Phase diagram of the $t$-$J$ model for various values of $J/t$.}
The spin exchange coupling constant is varied as $J/t=0.1, 0.3, 0.5, 1.0$ from ({\bf a}) to ({\bf d}).
This result is obtained at $N_s=16$.
}
\label{fig:J_variation}
\end{figure}

{\bf Phase diagram of the $t$-$J$ model with variation of $J/t$.}
Figure~\ref{fig:J_variation} shows how the phase diagram of the $t$-$J$ model changes with variation of $J/t$.
As one can see from Fig.~\ref{fig:J_variation}~({\bf a}), the optimal pairing amplitude producing the maximum NWOS is obtained along $\Delta=0$ for small $J/t$, meaning that there is no pairing correlation at this parameter regime.

The optimal pairing amplitude begins to be lifted from $\Delta=0$ beyond a critical value of $J/t$ roughly larger than 0.3. 
As $J/t$ increases further, the optimal pairing amplitude also gets higher, expanding the regime of superconductivity. 

Too large values of $J/t$ may not be physically meaningful in the single-band model, where the $t$-$J$ model is obtained as the large-$U$ expansion of the Hubbard model with $J/t$ being proportional to $t/U$~\cite{Gros87, MacDonald88}. 
It is, however, worthwhile to note that the spin exchange coupling can be also generated by the $d$-$p$ hybridization mechanism in the three-band model, where the value of $J/t$ can go beyond the $t/U$ scaling~\cite{Zhang88}

{\bf Construction of the RVB state.}
We begin with the usual form of the BCS state, which can be given in the momentum space as follows:
\begin{align}
|\psi_{\rm BCS}\rangle &= \prod_{\bf k} (u_{\bf k}+v_{\bf k}c^\dagger_{{\bf k}\uparrow}c^\dagger_{-{\bf k}\downarrow}) |0\rangle \nonumber \\
&= {\cal N} \prod_{\bf k} (1+g_{\bf k}c^\dagger_{{\bf k}\uparrow}c^\dagger_{-{\bf k}\downarrow}) |0\rangle \nonumber \\
&= {\cal N} e^{\sum_{\bf k} g_{\bf k}c^\dagger_{{\bf k}\uparrow}c^\dagger_{-{\bf k}\downarrow}}  |0\rangle \nonumber \\
&= {\cal N} e^{\sum_{i,j} \tilde{g}_{ij}c^\dagger_{i\uparrow}c^\dagger_{j\downarrow}}  |0\rangle \nonumber \\
&=  {\cal N} \sum_n \frac{1}{n!} \Big( \sum_{i,j} \tilde{g}_{ij}c^\dagger_{i\uparrow}c^\dagger_{j\downarrow} \Big)^n |0\rangle 
\end{align}
where $u_{\bf k}^2=\frac{1}{2}(1+\xi_{\bf k}/E_{\bf k})$ and $v_{\bf k}^2=\frac{1}{2}(1-\xi_{\bf k}/E_{\bf k})$ with $\xi_{\bf k}=\epsilon_{\bf k}-\mu$ and $E_{\bf k}=\sqrt{\xi_{\bf k}^2+\Delta_{\bf k}^2}$.
$\tilde{g}_{ij}=\tilde{g}({\bf r}_i - {\bf r}_j)$ is the Fourier transform of $g_{\bf k}=v_{\bf k}/u_{\bf k}$, and ${\cal N}$ is the normalization constant. 
Note that the second line is obtained under the condition that $u_{\bf k} \neq 0$ for all ${\bf k}$. 
We will address what happens if this condition is violated later.

The Gutzwiller projection can be conveniently implemented if one works with the real-space basis states, which are already in the projected Hilbert space with no double occupancy.
For example, consider that the spin up and down electrons are located in a given configuration set of $\{{{\bf r}_\uparrow}, {\bf r}_\downarrow\}$.
Then, the amplitude of the RVB state for this configuration set can be given as follows:
\begin{align}
{\cal A}_{\rm RVB}(\{{\bf r}_\uparrow,{\bf r}_\downarrow\})=\det{(\tilde{g}_{ij})}  ,
\label{eq:RVB_simple}
\end{align}
where $i$ and $j$ run though the coordinates of all spin up and down electrons, respectively~\cite{Gros88}. 

Now, let us come back to what happens if $u_{\bf k}= 0$ at some momenta. 
In this situation, which can occur for the $d$-wave pairing symmetry, the amplitude of the RVB state takes a rather complicated form.
Specifically, one has to take into account the contributions from $u_{\bf k} = 0$ separately from those from otherwise. 
For example, let us say that $u_{\bf k} = 0$ at ${\bf k}=\overline{\bf k}$.
Then, the plane wave states at $\overline{\bf k}$ and $-\overline{\bf k}$ are ``preoccupied'' by the spin up and down electrons, respectively, rather than forming Cooper pairs.  

Consequently, the amplitude of the RVB state should be modified as follows:
\begin{align}
{\cal A}_{\rm RVB}(\{{\bf r}_\uparrow,{\bf r}_\downarrow\})=
&\sum_{\forall \; {\rm comb.}}  
{(-1)^p} \det{(\tilde{g}_{ij})}  \nonumber \\
&\times \det{(e^{i\overline{\bf k}_l\cdot\overline{\bf r}_{m\uparrow}})} \det{(e^{-i\overline{\bf k}_l\cdot\overline{\bf r}_{n\downarrow}})},
\label{eq:RVB_direct}
\end{align}
where 
$\overline{\bf k}_l$ is the $l$-th preoccupied momentum, and
$\overline{\bf r}_{m\uparrow}$ and $\overline{\bf r}_{n\downarrow}$ are the coordinates of the $m$-th spin-up and $n$-th spin-down electrons, respectively, belonging to the preoccupied plane wave states. 
The sum is taken over all possible combinations of choosing $\{ \overline{\bf r}_\uparrow, \overline{\bf r}_\downarrow \}$ out of $\{ {\bf r}_\uparrow, {\bf r}_\downarrow \}$.
$p$ is the permutation parity occurring when all creation operators are rearranged in a predetermined convention.

Unfortunately, computing Eq.~\eqref{eq:RVB_direct} turns out to be quite time-consuming since there are simply too many different combinations. 
To reduce the computing time, we employ a trick of adding a very small constant to the pairing amplitude, i.e., $\Delta_{\bf k} \rightarrow \Delta_{\bf k}+\delta$, which eliminates the preoccupied momenta, leaving only the single combination, where all electrons are Cooper-paired.
At the end of computation, we take the limit of vanishing $\delta$.
In Figs.~\ref{fig:RVBmu} and \ref{fig:RVBx}, we take $\delta/t=0.0001$.

{\bf Acknowledgements} 

The authors are grateful to Jainendrra K. Jain, Sutirtha Mukherjee, and Yunkyu Bang for insightful discussions. 
The authors thank Center for Advanced Computation (CAC) at Korea Institute for Advanced Study for providing computing resources for this work.
K.P. acknowledges the hospitality of Subir Sachdev during his sabbatical visit to Harvard University, where parts of this work have been completed. 


\end{document}